\documentclass[preprint,12pt,authoryear]{elsarticle} 

\usepackage{bbm}
\usepackage{fullpage}
\usepackage{amsmath}
\usepackage{mathrsfs}
\usepackage{subfig}
\usepackage[hidelinks]{hyperref}
\hypersetup{colorlinks,bookmarksopen,linkcolor=blue,citecolor=blue,urlcolor=blue}

\newcommand{\bs}[1]{{\boldsymbol{#1}}}

\usepackage{amssymb}

\journal{arxiv}

\begin{document}
	
\begin{frontmatter}
		
		
		
\title{The response of grandstands driven by filtered Gaussian white noise processes}

		
\author[affiliation1]{Ond\v{r}ej Roko\v{s}\corref{mycorrespondingauthor}}
\cortext[mycorrespondingauthor]{Corresponding author.}
\ead{ondrej.rokos@fsv.cvut.cz}

\author[affiliation1]{Ji\v{r}\'{i} M\'{a}ca}
\ead{maca@fsv.cvut.cz}
		
\address[affiliation1]{Department of Mechanics, Faculty of Civil Engineering, Czech Technical University in Prague, Th\'{a}kurova~7, 166~29 Prague~6, Czech Republic.}
		
\begin{abstract}
This paper presents a semi-analytical estimate of the response of a grandstand occupied by an active crowd and by a passive crowd. Filtered Gaussian white noise processes are used to approximate the loading terms representing an active crowd. Lumped biodynamic models with a single degree of freedom are included to reflect passive spectators occupying the structure. The response is described in terms of the first two moments, employing the It\^{o} formula and the state augmentation method for the stationary time domain solution. The quality of the approximation is compared on the basis of three examples of varying complexity using Monte Carlo simulation based on a synthetic generator available in the literature. For comparative purposes, there is also a brief review of frequency domain estimates.
\end{abstract}
		
		
\end{frontmatter}
	
	

\section{Introduction}
\label{sectintroduction}
The response of a grandstand can be resolved quite easily by linear dynamics methods, if we neglect all the randomness of the system. However, for a more accurate description, at least the most significant uncertainties need to be taken into account. The main uncertainties include
\begin{itemize}
	\item forcing terms resulting from active crowd movements, especially synchronized jumping,
	\item the uncertainties of the parameters in discrete biodynamic models---randomness of stiffness, mass and damping matrices,
	\item the size and spatial distribution of an active crowd and a passive crowd.
\end{itemize}
Further generalizations can take into account various kinds of nonlinearities, e.g. geometrical and material non-linearities and non-linearities of biodynamic models \citep{Huang}. However, the following restrictions will be assumed for the purposes of this paper: the material parameters of the structure are treated as deterministic, since their influence is negligible in comparison with the sources listed above and the scope of the overall response; the spatial distribution of the crowd is fixed; biodynamic models of the passive crowd are treated as deterministic.

The results of measurements carried out on simple structures indicate that an active spectator might be represented as a time dependent force, \textit{cf} \citep{Ellis}. In other words, under certain circumstances, an active spectator does not influence the properties of the structure. Several generators of normalized artificial load processes have been developed in the literature, e.g. \citep{Sim} and \citep{Racic}, which can be supplemented by human body weights, e.g. \citep{Hermanussen}. However, the passive part of the crowd is assumed to be stationary in space and in permanent contact with the structure, excluding accelerations that exceed the gravity of the earth. Thus a passive spectator can be modeled as a biodynamic system. A survey of models of this type can be found in \citep{Sachse}. These remarks allow us to model the structure considered here in terms of random vibrations of linear systems.

Monte Carlo simulation (MC) is one of the ways used in advanced grandstand design procedures for reflecting all kinds of randomness. MC is a general method, but it has considerable disadvantages in its original form, e.g. rather slow convergence when estimating low probabilities, and high computational demands. MC will be employed in this paper mainly for controlling the accuracy and the performance of semi-analytical methods introduced in terms of stochastic differential equations. These simplified methods can be applicable and useful in the preliminary stages of design procedures, when quick and approximate solutions are sufficient.

\section{Stochastic differential equations}
\label{SDE}
Employing the Finite Element Method, and taking into account the above-mentioned assumptions, a mathematical model of the structural system considered here can be written as a set of hyperbolic differential equations
\begin{equation}
\bs{M}\bs{\ddot{Z}}(t)+\bs{C}\bs{\dot{Z}}(t)+\bs{K}\bs{Z}(t)=\bs{GY}(t),\quad t\geq 0
\label{2eq1}
\end{equation}
where $\bs{Z}$ and $\bs{Y}$ are $\mathbb{R}^{d/2}$ and $\mathbb{R}^{d'}$-valued stochastic processes, and $\bs{M}$, $\bs{C}$, $\bs{K}$ and $\bs{G}$ are $(d/2,d/2)$ and $(d/2,d')$ matrices of mass, damping, stiffness and input distribution. Over-dot denotes a derivative with respect to time $\dot{g}(t)=d/dt\,g(t)$. To simplify subsequent expressions, let us apply expectation operator $\mathsf{E}$ in equation (\ref{2eq1}) and substract the result from (\ref{2eq1}). We arrive at the set of two equations
\begin{eqnarray}
&&\bs{M}\bs{\ddot{\mu}}_Z(t)+\bs{C}\bs{\dot{\mu}}_Z(t)+\bs{K}\bs{\mu}_Z(t)=\bs{G\mu}_Y(t),\quad t\geq 0\label{2eq2}\\
&&\bs{M}\bs{\ddot{\tilde{Z}}}(t)+\bs{C}\bs{\dot{\tilde{Z}}}(t)+\bs{K}\bs{\tilde{Z}}(t)=\bs{G\tilde{Y}}(t),\quad t\geq 0\label{2eq3}
\end{eqnarray}
for the mean value $\bs{\mu}_Z(t)=\mathsf{E}\bs{Z}(t)$ and centered process $\bs{\tilde{Z}}(t)=\bs{Z}(t)-\bs{\mu}_Z(t)$, $\bs{\mu}_Y(t)=\mathsf{E}\bs{Y}(t)$ and $\bs{\tilde{Y}}(t)=\bs{Y}(t)-\bs{\mu}_Y(t)$. As will become apparent in section \ref{appToGrandstands}, under certain conditions process $\bs{\tilde{Z}}(t)$ is approximately normal, and since differential equation (\ref{2eq3}) is linear with deterministic coefficients, it is reasonable to accept Gaussian approximation also for $\bs{\tilde{Y}}(t)$. This consideration leads us to stochastic differential equations and to the It\^{o} calculus. Under Gaussian assumptions, the response will be completely specified by its mean $\bs{\mu}_Z(t)$ and covariance $\bs{c}_Z(t,s)=\mathsf{E}[\bs{\tilde{Z}}(t)\bs{\tilde{Z}}(s)^T]$. The required quantities can be obtained from the time domain or from the frequency domain.

\subsection{Solution in the time domain}
\label{timedomain}
Equation (\ref{2eq2}) can be solved by direct integration or, more conveniently, by a Fourier series (or Fourier transform) assuming periodic mean $\bs{\mu}_Y(t)$, \textit{cf} section \ref{frequencydomain}. Let us rewrite (\ref{2eq3}) as
\begin{equation}
\frac{d}{dt}
\left[\begin{array}{c}
\bs{\tilde{Z}}(t)\\
\bs{\dot{\tilde{Z}}}(t)
\end{array}\right]
=
\left[\begin{array}{c c}
\bs{0} & \bs{I} \\
-\bs{M}^{-1}\bs{K} & -\bs{M}^{-1}\bs{C}
\end{array}\right]
\left[\begin{array}{c}
\bs{\tilde{Z}}(t)\\
\bs{\dot{\tilde{Z}}}(t)
\end{array}\right]
+
\left[\begin{array}{c}
\bs{0}\\
\bs{M}^{-1}\bs{G}
\end{array}\right]\bs{\tilde{Y}}(t)
\label{2eq4}
\end{equation}
or, in a more compact form,
\begin{equation}
\bs{\dot{\tilde{X}}}(t)=\bs{a}\bs{\tilde{X}}(t)+\bs{b}\bs{\tilde{Y}}(t),\quad t\geq 0
\label{2eq5}
\end{equation}
where $\bs{\tilde{X}}$ is an $\mathbb{R}^d$-valued state-space vector stochastic process with zero mean, and $\bs{a}$ and $\bs{b}$ are $(d,d)$ and $(d,d')$-matrices. The solution of this differential equation is given in the form
\begin{equation}
\bs{\tilde{X}}(t)=\bs{\theta}(t)\bs{\tilde{X}}(0)+\int_0^t\bs{\theta}(t-s)\bs{b}\bs{\tilde{Y}}(s)\,ds
\label{2eq6}
\end{equation}
where $\bs{\theta}(t-s)$ denotes a Green function or the unit impulse response satisfying
\begin{equation}
\frac{\partial\bs{\theta}(t-s)}{\partial t}=\bs{a\theta}(t-s),\quad t\geq s\geq 0,
\label{2eq7}
\end{equation}
$\bs{\theta}(0)=\bs{I}$ the identity and $\bs{\theta}(t-s)=\mbox{exp}[\bs{a}(t-s)]$ can be expressed as a matrix exponential, \textit{cf} \citep{Soong}. Initial conditions $\bs{\tilde{X}}(0)$ will be set to zero for simplicity. Forcing term $\bs{\tilde{Y}}(t)$ can also satisfy its own stochastic differential equation driven by Gaussian white noise $\bs{W}(t)=d\bs{B}(t)/dt$. For example, let $\hat{Y}_1(t)$ be a continuous-time Gaussian auto-regression scalar process of order $p$, denoted as $AR(p)$, \textit{cf} \citep{Brockwell}. Then $\hat{Y}_1(t)=\bs{e}_1^T\bs{S}_1(t)$ where the state vector $\bs{S}_1(t)=[S_{1,1}(t),\ldots,S_{1,p}(t)]^T$ satisfies the It\^{o} equation
\begin{equation}
d\bs{S}_1(t)=\bs{A}_1\bs{S}_1(t)dt+\bs{b}_1dB(t),
\label{2eq8}
\end{equation}
\begin{equation*}
\bs{A}_1=\left[\begin{array}{c c c c c}
0 & 1 & 0 & \ldots & 0 \\
0 & 0 & 1 & \ldots & 0 \\
\vdots & \vdots & \vdots & \ddots & \vdots \\
0 & 0 & 0 & \ldots & 1 \\
-a_p & -a_{p-1} & -a_{p-2} & \ldots & -a_1
\end{array}\right]
\mbox{,}\quad
\bs{e}_1=\left[\begin{array}{c}
1 \\ 0 \\ \vdots \\ 0 \\ 0
\end{array}\right]
\quad\mbox{and}\quad
\bs{b}_1=\left[\begin{array}{c}
0 \\ 0 \\ \vdots \\ 0 \\ a_0
\end{array}\right].
\end{equation*}
Processes of this kind are also called filtered white noise processes or colored processes, and they have a specific frequency content. Let us assume that $\tilde{Y}_i(t)$ of  $\bs{\tilde{Y}}(t)=[\tilde{Y}_1(t),\ldots,\tilde{Y}_{d'}(t)]^T$ are mutually independent $AR(p_i)$ processes. Then we can merge equations (\ref{2eq4}) and (\ref{2eq8}) to obtain one coupled system
\begin{eqnarray}
&d\left[\begin{array}{c}
\bs{\tilde{Z}}(t)\\
\bs{\dot{\tilde{Z}}}(t)\\
\bs{S}_1(t)\\
\vdots\\
\bs{S}_{d'}(t)
\end{array}\right]
=
\left[\begin{array}{c c c c c}
\bs{0} & \bs{I} & \bs{0} & \ldots & \bs{0} \\
-\bs{M}^{-1}\bs{K} & -\bs{M}^{-1}\bs{C} & \bs{M}^{-1}\bs{G}\bs{d}_1\bs{e}_1^T & \ldots & \bs{M}^{-1}\bs{G}\bs{d}_{d'}\bs{e}_{d'}^T \\
\bs{0} & \bs{0} & \bs{A}_1 & \ldots & \bs{0} \\
\bs{0} & \bs{0} & \bs{0} & \ddots & \bs{0} \\
\bs{0} & \bs{0} & \bs{0} & \ldots & \bs{A}_{d'} \\
\end{array}\right]&
\left[\begin{array}{c}
\bs{\tilde{Z}}(t)\\
\bs{\dot{\tilde{Z}}}(t)\\
\bs{S}_1(t)\\
\vdots\\
\bs{S}_{d'}(t)
\end{array}\right]dt+\nonumber\\
&+
\left[\begin{array}{c c c}
\bs{0} & \ldots & \bs{0} \\
\bs{0} & \ldots & \bs{0} \\
\bs{b}_1 & \ldots & \bs{0} \\
\bs{0} & \ddots & \bs{0} \\
\bs{0} & \ldots & \bs{b}_{d'} \\
\end{array}\right]d\bs{B}(t)&
\label{2eq9}
\end{eqnarray}
where $\bs{d}_i$ are column vectors with the unit in $i$-th position, and $\bs{B}(t)$ is an $\mathbb{R}^{d'}$-valued Brownian motion. This approach is called a state augmentation method \citep{Grigoriu_stoch}. An extension to the case $\tilde{Y}_i(t)=\sum_{k=1}^n\hat{Y}_k(t)$, where $\hat{Y}_k(t)$ are mutually independent $AR(p)$ processes, is carried out in an obvious manner. This methodology will be employed in section \ref{appToGrandstands} for $AR(2)$ processes. Equation (\ref{2eq9}) can again be rewritten in compact form
\begin{equation}
d\bs{X}(t)=\bs{a}\bs{X}(t)dt+\bs{b}d\bs{B}(t),\quad t\geq 0,
\label{2eq10}
\end{equation}
and employing the It\^{o} formula for semimartingales we arrive at the system of evolutionary equations for the response mean $\bs{\mu}_X(t)$ and covariance $\bs{c}_X(t,s)$
\begin{eqnarray}
\bs{\dot{\mu}}_X(t)&=&\bs{a\mu}_X(t),\quad t\geq0,\label{2eq11}\\
\bs{\dot{c}}_X(t,t)&=&\bs{ac}_X(t,t)+\bs{c}_X(t,t)\bs{a}^T+\bs{bb}^T,\quad t\geq0,\label{2eq12}\\
\frac{\partial\bs{c}_X(t,s)}{\partial t}&=&\bs{ac}_X(t,s),\quad t>s\geq0.\label{2eq13}
\end{eqnarray}
Since the driving forces $d\bs{B}(t)$ are Gaussian white noise and the coefficients are constant in time, the solution is an Ornstein-Uhlenbeck process with an existing stationary solution. In our case, stationary mean $\bs{\mu}_X=\bs{0}$ and covariance $\bs{\dot{c}}_X(t,t)=\bs{\dot{c}}_X(t-t)=\bs{\dot{c}}_X=\bs{0}$ which leads to the so-called continuous Lyapunov equation
\begin{equation}
\bs{0}=\bs{ac}_X+\bs{c}_X\bs{a}^T+\bs{bb}^T.
\label{2eq14}
\end{equation}
For details and further developments, see \citep{Grigoriu_stoch}. Since the stationary matrix $\bs{c}_X$ contains only response displacements and velocities, the variances of the acceleration have to be computed through the following formulas which are valid for weakly stationary processes
\begin{equation}
\bs{c}_{\dot{X}}=\left.-\frac{d^2\bs{c}_X(t)}{dt^2}\right|_{t=0}=-\bs{a}^2\bs{c}_X,
\label{2eq14a}
\end{equation}
where $\bs{a}^2$ denotes matrix power and $\bs{c}_{\dot{X}}$ denotes the stationary covariance matrix of velocities and accelerations. Equation (\ref{2eq14a}) is evaluated employing (\ref{2eq13}), which in our special case is simplified to
\begin{equation}
\bs{c}_X(t)=\bs{\theta}(t)\bs{c}_X=\exp[\bs{a}t]\bs{c}_X.
\label{2eq14b}
\end{equation}
%
\subsection{Solution in the frequency domain}
\label{frequencydomain}
Taking the Fourier transform of equation (\ref{2eq2}) leads to
\begin{equation}
\bs{\hat{\mu}}_Z(\omega)=\bs{H}(\omega)\bs{G\hat{\mu}}_Y(\omega)
\label{2eq15}
\end{equation}
where the FRF (Frequency Response Function) $\bs{H}(\omega)$ is
\begin{equation}
\bs{H}(\omega)=[-\omega^2\bs{M}+\mathbbm{i}\omega\bs{C}+\bs{K}]^{-1},
\label{2eq16}
\end{equation}
$\mathbbm{i}$ denotes a complex unit, $\omega$ denotes angular frequency and $\hat{g}(\omega)$ denotes the Fourier transform of function $g(t)$. Assuming $\bs{\mu}_Y(t)$ in periodic form $\bs{\hat{\mu}}_Y(\omega)=\sum_{k=1}^{n_{\mathrm{harm}}}\sqrt{2\pi}/2\,\bs{r}_k[\exp\{\mathbbm{i}\varphi_k\}\delta(\omega-\bar{\omega}_k)+\exp\{-\mathbbm{i}\varphi_k\}\delta(\omega+\bar{\omega}_k)]$ where $\bs{r}_k$, $\varphi_k$ and $\bar{\omega}_k$ are amplitude, phase shift and angular frequency of the $k$-th harmonic, we obtain the mean response as a solution of the complex linear system of equations.

Employing spectral decomposition of stationary random processes, the response variances are acquired in terms of spectral density matrices $\bs{S}_{\tilde{Y}\tilde{Y}}$ and $\bs{S}_{\tilde{Z}\tilde{Z}}$, $(S_{\tilde{Y}\tilde{Y}}(\omega))_{ii}=\hat{f}_{\tilde{Y}}(\omega)$, where $\hat{f}_{\tilde{Y}}(\omega)$ is a spectral density estimate of the centered forcing term
\begin{equation}
\hat{f}_{\tilde{Y}}(\omega)=\mathsf{E}\int_{-\infty}^\infty b(x-\omega)I_T(\omega)\,dx,
\label{2eq17}
\end{equation}
$b(x)$ is some weight function, \textit{cf} \citep{Andel}, and $I_{T}(\omega)$ denotes the corresponding periodogram
\begin{equation}
I_T(\omega)=\frac{1}{2\pi T}\left|\int_0^T \tilde{Y}(t)e^{-\mathbbm{i}t\omega}\,dt\right|^2, \qquad -\infty<\omega<\infty.
\label{2eq18}
\end{equation}
The diagonal form of $\bs{S}_{\tilde{Y}\tilde{Y}}$ suggests that we treat all input processes as independent. Knowing the spectral density matrix of the input vector stochastic process, we obtain the spectral density matrix of the output process according to \cite{Soong}
\begin{equation}
\bs{S}_{\tilde{Z}\tilde{Z}}(\omega)=\bs{H}(\omega)\bs{GS}_{\tilde{Y}\tilde{Y}}(\omega)\bs{G}^T\bs{H}^{\dagger}(\omega)
\label{2eq19}
\end{equation}
where $\bs{H}^{\dagger}(\omega)$ denotes a Hermitian transpose to $\bs{H}(\omega)$. The variance of the stationary scalar process $\tilde{Z}(t)$ with two-sided spectral density $f_{\tilde{Z}}(\omega)$ or with one-sided spectral density $g_{\tilde{Z}}(\omega)$ is evaluated as
\begin{equation}
\sigma_{\tilde{Z}}^2=\int_{-\infty}^\infty f_{\tilde{Z}}(\omega)\ d\omega=\int_{0}^\infty g_{\tilde{Z}}(\omega)\,d\omega
\label{2eq20}
\end{equation}
and the variance of time derivative $\dot{\tilde{Z}}(t)$
\begin{equation}
\sigma_{\dot{{\tilde{Z}}}}^2=\dot{\sigma}_{\tilde{Z}}^2=\int_{-\infty}^\infty \omega^2f_{\tilde{Z}}(\omega)\,d\omega=\int_{0}^\infty \omega^2g_{\tilde{Z}}(\omega)\,d\omega.
\label{2eq21}
\end{equation}
By analogy for higher time derivatives, applying higher powers of angular frequency $\omega$.
%

\subsection{Transformation to modal coordinates}
\label{modalsol}
It is desirable to reduce the system unknowns in equations (\ref{2eq1}), (\ref{2eq2}), (\ref{2eq14}) and (\ref{2eq19}) for MC, mean value, time domain and frequency domain solutions. The usual modal transform technique can be employed directly when a structure with proportional damping is loaded by active spectators only. Some problems arise when the passive part of the crowd is introduced. All matrices are extended according to the added degrees of freedom (dofs). Biodynamic models possess non-proportional damping, so complex eigenvectors should be used. Simultaneously, all these models have the first eigenfrequency close to 5 Hz (and the second eigenfrequency close to 8 Hz if using two-degrees-of-freedom models). This implies that the number of eigenvectors employed in the transform is considerably enlarged. Another possibility is partial modal transformation. The system matrices can be decomposed into four parts according to
\begin{equation}
\bs{A}=\left(\begin{array}{cc}
\bs{A}_{SS} & \bs{A}_{SH} \\
\bs{A}_{HS} & \bs{A}_{HH} \\
\end{array}\right)
\label{modal1}
\end{equation}
where $\bs{A}$ stands either for mass, stiffness or damping matrix. Sub-matrix $\bs{A}_{SS}$ corresponds to the structure, $\bs{A}_{HH}$ to the passive crowd, and $\bs{A}_{SH}$ or $\bs{A}_{HS}$ represents the mutual interactions. Let us note that sub-matrix $\bs{A}_{HH}$ is a diagonal or a band, and $\bs{A}_{SS}$ is sparse. Computing the eigenvectors corresponding to an empty structure, i.e. corresponding to $\bs{A}_{SS}$, arranged in $\bs{\tilde{V}}$ and writing
\begin{equation}
\bs{V}=\left(\begin{array}{cc}
\bs{\tilde{V}} & \bs{0} \\
\bs{0} & \bs{I} \\
\end{array}\right),
\label{modal2}
\end{equation}
all equations can be partially transformed by $\bs{V}$, as in the standard procedure. Then the dofs inherent to a passive crowd are unchanged, but the structure response is described through several modal coordinates.

Concerning the Lyapunov equation (\ref{2eq14}) partially transformed into modal coordinates, we can further reduce the computational effort by some prior information. Let us assume that the system response corresponds to the equation (\ref{2eq9}) with $p_i=2$. Splitting all matrices in (\ref{2eq14}) leads to
\begin{equation}
\left(\begin{array}{cc}
\bs{a}_{11} & \bs{a}_{12} \\
\bs{a}_{21} & \bs{a}_{22} \\
\end{array}\right)
\left(\begin{array}{cc}
\bs{c}_{11} & \bs{c}_{12} \\
\bs{c}_{21} & \bs{c}_{22} \\
\end{array}\right)
+
\left(\begin{array}{cc}
\bs{c}_{11} & \bs{c}_{12} \\
\bs{c}_{21} & \bs{c}_{22} \\
\end{array}\right)
\left(\begin{array}{cc}
\bs{a}^T_{11} & \bs{a}^T_{21} \\
\bs{a}^T_{12} & \bs{a}^T_{22} \\
\end{array}\right)
+
\left(\begin{array}{cc}
\bs{0} & \bs{0} \\
\bs{0} & \bs{Q}_{22} \\
\end{array}\right)
=\bs{0},
\label{modal3}
\end{equation}
where
\begin{equation}
\bs{a}_{11}=\left(\begin{array}{cc}
\bs{0} & \bs{I} \\
-[\bs{V}^T\bs{M}\bs{V}]^{-1}[\bs{V}^T\bs{K}\bs{V}] & -[\bs{V}^T\bs{M}\bs{V}]^{-1}[\bs{V}^T\bs{C}\bs{V}] \\
\end{array}\right),
\label{modal4}
\end{equation}
$\bs{a}_{21}=\bs{0}$ and remaining sub-matrices have obvious structure. Dropped subscript at covariance matrix $\bs{c}$ emphasizes partial transformation to modal coordinates. Since $\bs{bb}^T$ resp. $\bs{Q}_{22}$ are symmetric, in fact diagonal, the solution will be also symmetric, $\bs{c}_{12}=\bs{c}^T_{21}$. In the case of $AR(2)$ processes, sub-matrix $\bs{c}_{22}$ can be computed explicitly; single $AR(2)$ process has uncorrelated state variables $S_1$, $S_2$ and $\mathsf{var}S_1=a_0^2/(2a_1a_2)$, $\mathsf{var}S_2=a_0^2/(2a_1)$, based on moment equations, thus $\bs{c}_{22}$ is a diagonal matrix. Introduced considerations reduce the system of four equations (\ref{modal3}) in expanded form, to the set of two coupled equations
\begin{eqnarray}
\bs{a}_{11}\bs{c}_{12}+\bs{c}_{12}\bs{a}^T_{22}+\bs{a}_{12}\bs{c}_{22} &=& \bs{0} \label{modal5}\\
\bs{a}_{11}\bs{c}_{11}+\bs{c}_{11}\bs{a}^T_{11}+\bs{c}_{12}\bs{a}^T_{12} + \bs{a}_{12}\bs{c}^T_{12} &=& \bs{0}\label{modal6}
\end{eqnarray}
for unknowns $\bs{c}_{12}$ and $\bs{c}_{11}$. The set resembles Sylvester and Lyapunov equations respectively with reduced size. Backward transformation $\bs{\bar{c}}_X=\bs{V\bar{c}}\bs{V}^T$ gives covariance matrix for displacement or velocity vector, here $\bs{\bar{c}}$ denotes an appropriate sub-matrix of $\bs{c}$ storing the modal displacements or velocities, $\bs{\bar{c}}_X$ then contains the nodal displacements or velocities.
%
\subsection{Crossings of Gaussian processes}
\label{upcrossing}
One of the measures of system performance is level crossing. Under some circumstances, in a stationary case, it can be shown that up-crossing is directly connected with the reliability of the system. The $x$-up-crossing rate of Gaussian process $X(t)$ with non-stationary mean value $\mu(t)$ and stationary variance $\sigma^2$ is estimated as \citep{Soong}
\begin{equation}
\nu_x^+(t)=\frac{\dot{\sigma}}{\sigma}\left[\phi\left(\frac{\dot{\mu}(t)}{\dot{\sigma}}\right)+\frac{\dot{\mu}(t)}{\dot{\sigma}}\Phi\left(\frac{\dot{\mu}(t)}{\dot{\sigma}}\right)\right]\phi\left(\frac{x-\mu(t)}{\sigma}\right),
\label{2eq22}
\end{equation}
where $\nu_x^+(t)$ is the $x$-up-crossing rate of level $x$ at time $t$, $\phi(\alpha)=1/\sqrt{2\pi}\exp{-\alpha^2/2}$, $\Phi(u)=\int_{-\infty}^u\phi(\alpha)\,d\alpha$, $\sigma^2=\mathsf{var}X(t)$, $\dot{\sigma}^2=\mathsf{var}\dot{X}(t)$. The total mean number of upcrossings in time interval $[0,T]$ is computed according to
\begin{equation}
n_x^+(T)=\int_0^T\nu_x^+(t)\,dt.
\label{2eq23}
\end{equation}
The relations can be generalized to $D$-out-crossings of a $d$-valued stochastic process, where $D$ is some set in $\mathbb{R}^d$.

Another measure of system performance from the point of view of serviceability is the root mean square value estimated as
\begin{equation}
RMS=\sqrt{\frac{1}{T}\int_0^TX(t)^2\,dt}=\sqrt{\frac{1}{T}\int_0^T\mu_X(t)^2\,dt+\sigma^2},
\label{2eq24}
\end{equation}
where $\sigma^2$ denotes the stationary variance of centered process $\tilde{X}(t)$, and $\mu_X(t)$ its mean value. Analogous formulas are valid for velocity and acceleration.
%
\section{Applications to the response of grandstands}
\label{appToGrandstands}
\begin{figure}
	\centering
  \subfloat[]{\includegraphics[scale=0.7]{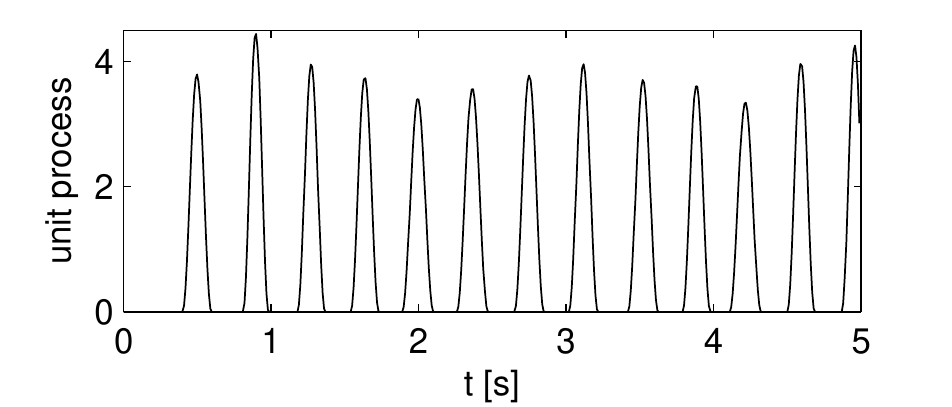}}
  \subfloat[]{\includegraphics[scale=0.7]{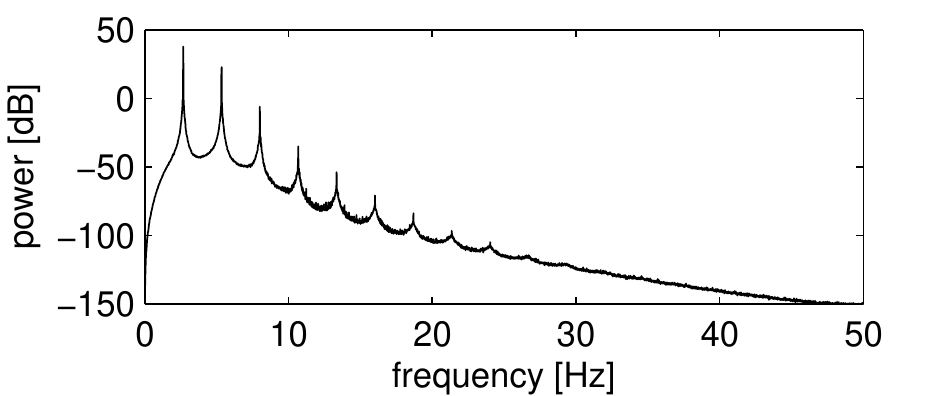}}
	\caption{Single time history (a) and power spectral density of 10~000 realizations (b) of forcing term $Y(t)$ generated according to \cite{Sim}.}
	\label{3fig1}
\end{figure}

\begin{figure}
	\centering
  \subfloat[]{\includegraphics[scale=0.7]{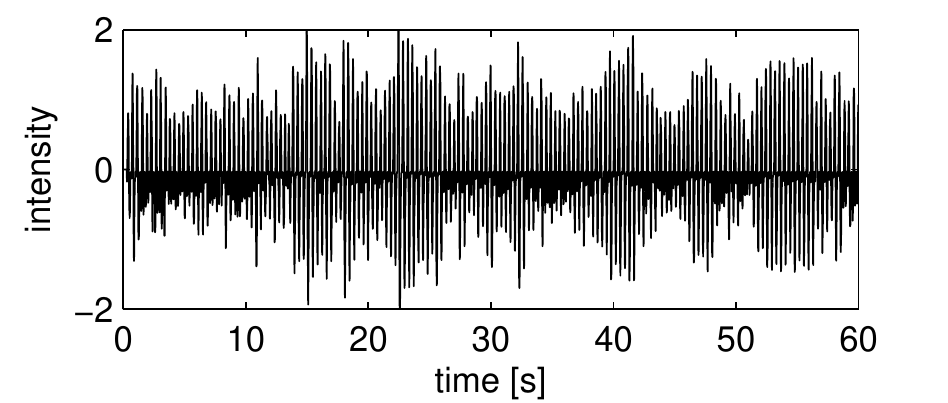}}
  \subfloat[]{\includegraphics[scale=0.7]{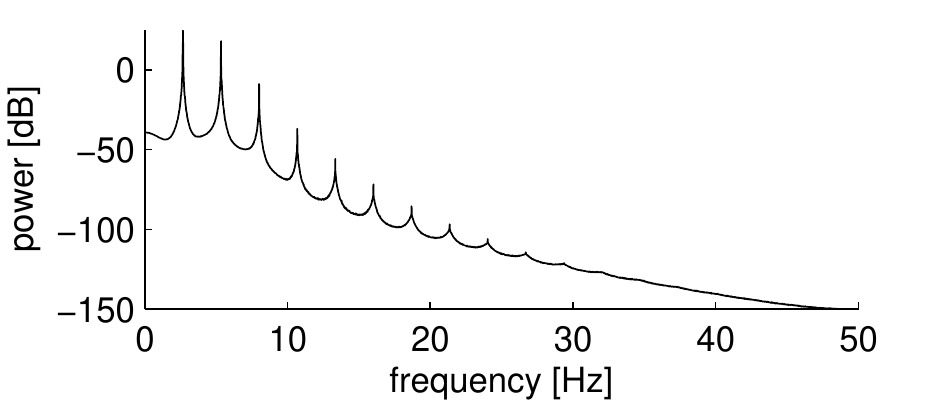}}\\
  \subfloat[]{\includegraphics[scale=0.7]{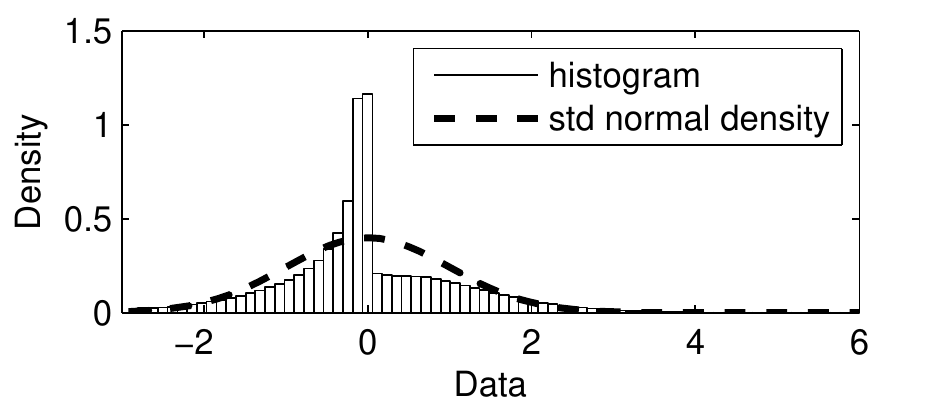}}
  \subfloat[]{\includegraphics[scale=0.7]{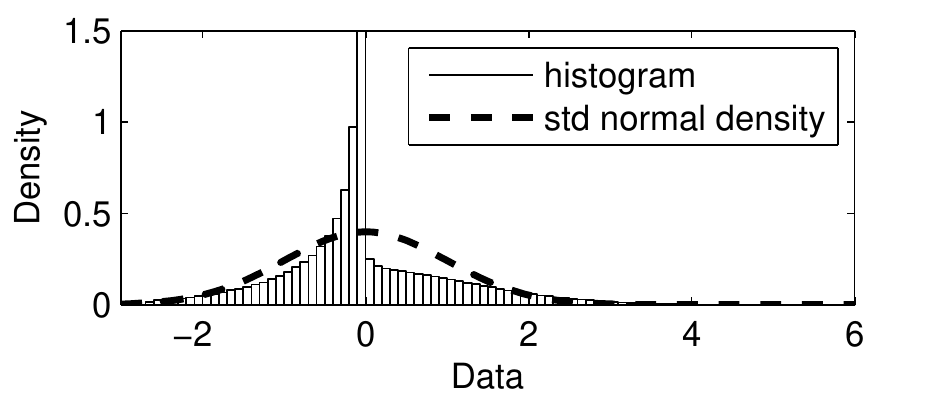}}
	\caption{Single centered time history of $\tilde{Y}(t)$ (a), the corresponding power spectral density of 10~000 realizations (b), a normalized histogram with standard normal density (c) and a normalized histogram of a non-unit process scaled with $G_H\mbox{[kN]}\sim \mathcal{N}(0.7709,0.0167)$ (d).}
	\label{3fig2}
\end{figure}

As was noted in section \ref{sectintroduction}, an active spectator can be treated as a time-dependent process. Figure \ref{3fig1} shows a single realization and the spectral density of a unit process, i.e. of the process with $G_H=1$, where $G_H$ denotes the weight of a spectator, jumping frequency $\bar{f}=2.67$ Hz. Spectral densities were computed from (\ref{2eq17}) with Parzen weight. The realization was generated according to \cite{Sim}.
Since this function is highly periodic, we will search the mean value in the form
\begin{equation}
\mu_Y(t)=\alpha_0+\sum_{k=1}^p\alpha_k\cos(k\cdot2\pi\bar{f}t)+\beta_k\sin(k\cdot2\pi\bar{f}t).
\label{3eq1}
\end{equation}
Then vector $\bs{\hat{\alpha}}$ of the estimated parameters $\hat{\alpha}_0,\hat{\alpha}_1,\dots,\hat{\alpha}_p,\hat{\beta}_1,\dots,\hat{\beta}_p$ can be found by the linear Least Squares Method as
\begin{equation}
\bs{\hat{\alpha}}=(\bs{\Phi}^T\bs{\Phi})^{-1}\bs{\Phi}^T{\bs{\bar{\mu}}_Y},
\label{3eq2}
\end{equation}
where 
\begin{equation*}
\bs{\Phi}=
\left(\begin{array}{c c c c c c}
1 & \cos(2\pi\bar{f}t_1) & \sin(2\pi\bar{f}t_1) & \dots & \cos(p\cdot2\pi\bar{f}t_1) & \sin(p\cdot2\pi\bar{f}t_1)\\
1 & \cos(2\pi\bar{f}t_2) & \sin(2\pi\bar{f}t_2) & \dots & \cos(p\cdot2\pi\bar{f}t_2) & \sin(p\cdot2\pi\bar{f}t_2)\\
\vdots & \vdots & \vdots & \ddots & \vdots & \vdots\\
1 & \cos(2\pi\bar{f}t_n) & \sin(2\pi\bar{f}t_n) & \dots & \cos(p\cdot2\pi\bar{f}t_n) & \sin(p\cdot2\pi\bar{f}t_n)\\
\end{array}\right),
\end{equation*}
$t_1,\dots,t_n$ is a fine enough and equidistant partition of the time interval, ${\bs{\bar{\mu}}_Y}=[\bar{\mu}_Y(t_1),\dots,\bar{\mu}_Y(t_n)]^T$ with
\begin{equation*}
\bar{\mu}_Y(t_i)=\frac{1}{N}\sum_{k=1}^NY_k(t_i),
\end{equation*}
are means over $N$ realizations $Y_k(t_i)$ in time instants $t_i$. Generating 10 000 trajectories provides the coefficients given in tab. \ref{3tab1}.
\begin{table}
	\centering
	\caption{Coefficients $\bs{\hat{\alpha}}$ for an approximation of the mean value in equation (\ref{3eq1}) for $\bar{f}=2.67$ Hz.}
	\begin{tabular}{|c|c|c|c|}\hline
	$\hat{\alpha}_0$ & \multicolumn{3}{c|}{0.9958} \\\hline
	$\hat{\alpha}_1$ & 0.2939 & $\hat{\beta}_1$ & 1.1170 \\
	$\hat{\alpha}_2$ & -0.2471 & $\hat{\beta}_2$ & 0.0984 \\
	$\hat{\alpha}_3$ & -0.0037 & $\hat{\beta}_3$ & -0.0153 \\
	$\hat{\alpha}_4$ & -0.0008 & $\hat{\beta}_4$ & -0.0001 \\\hline
	\end{tabular}
	\label{3tab1}
\end{table}
Realization centered with the mean value according to equation (\ref{3eq1}) and the coefficients from table \ref{3tab1} is  depicted in figure \ref{3fig2}, together with the spectral density and a normalized histogram, i.e. a histogram of the process $\bar{Y}(t)=[Y(t)-\mu(t)]/\sigma_Y=\tilde{Y}(t)/\sigma_Y$ where $\sigma_Y=\sqrt{\mathsf{var}Y(t)}=\sqrt{0.7627}$ is a stationary standard deviation.
The centered process resembles non-Gaussian colored noise. To be more precise, for $\bar{Y}(t)$ we have $\mu_{\bar{Y}}=0$ for the mean, $\mathsf{var}\bar{Y}=1$ for the variance, $\gamma_{3,\bar{Y}}=0.424$ for the coefficient of skewness, and $\gamma_{4,\bar{Y}}=4.076$ for the coefficient of kurtosis. For comparison, the standard Gaussian process has coefficients $0$, $1$, $0$ and $3$.

\begin{figure}
	\centering
  \includegraphics[scale=0.7]{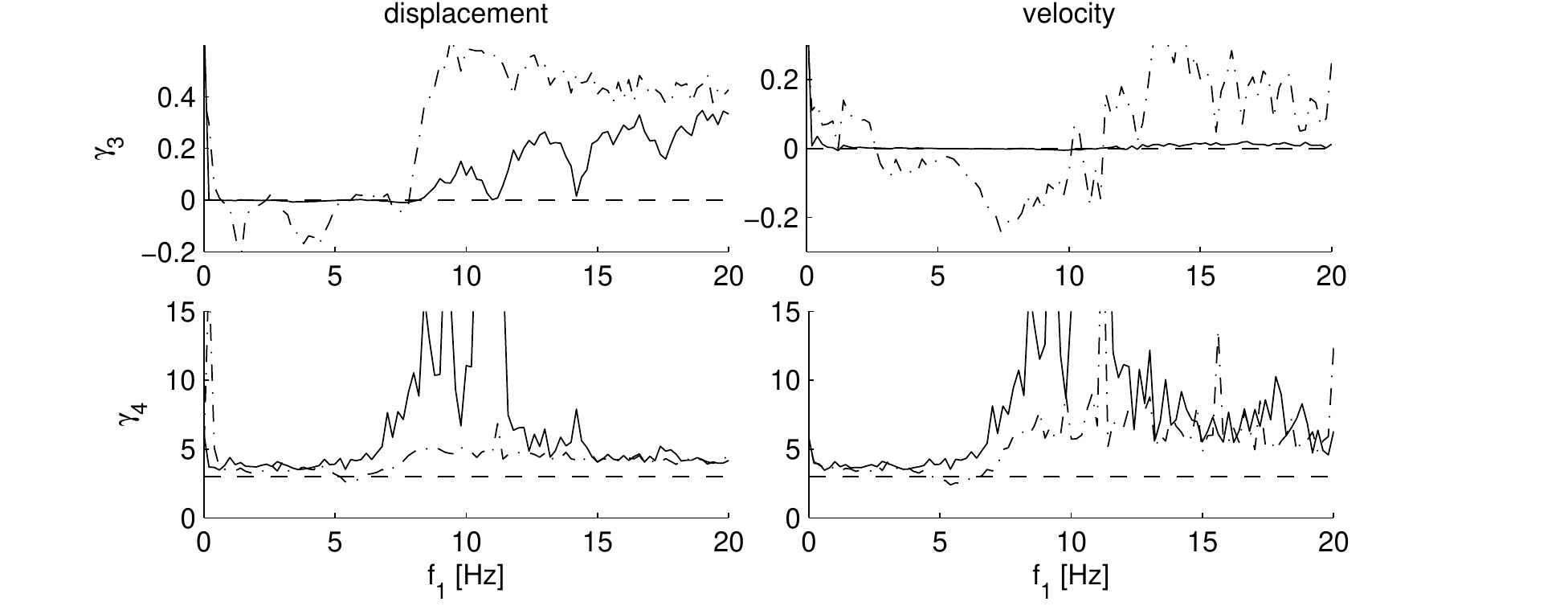}
	\caption{Coefficient of skewness $\gamma_3$ and coefficient of kurtosis $\gamma_4$ for the normalized displacement and velocity of the harmonic oscillator forced by the jumping process as functions of eigenfrequency $\mbox{f}_1$. Solid line - viscous damping $\zeta=0.001$; dash-dot line - $\zeta=0.07$; dashed line - corresponding values for the Gaussian random variable.}
	\label{3fig3}
\end{figure}

\begin{figure}
	\centering
  \subfloat[$\mbox{f}_1=4$ Hz, $\zeta=0.001$]{\includegraphics[scale=0.7]{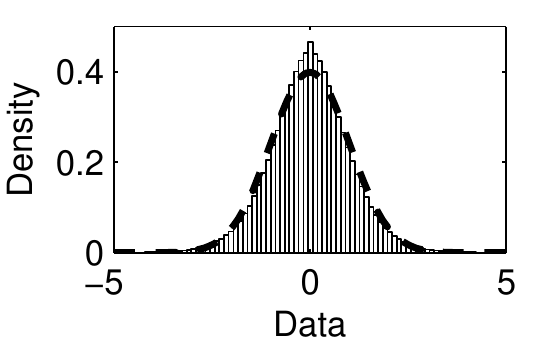}}
  \subfloat[$\mbox{f}_1=12$ Hz, $\zeta=0.001$]{\includegraphics[scale=0.7]{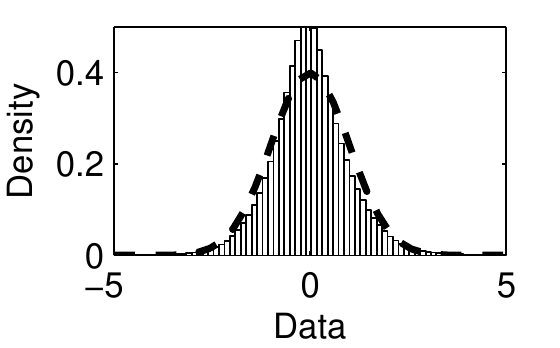}}\\
  \subfloat[$\mbox{f}_1=4$ Hz, $\zeta=0.07$]{\includegraphics[scale=0.7]{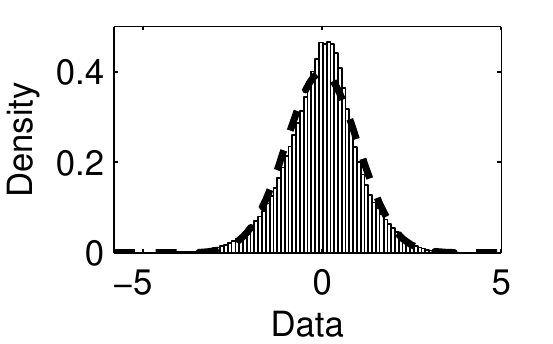}}
  \subfloat[$\mbox{f}_1=12$ Hz, $\zeta=0.07$]{\includegraphics[scale=0.7]{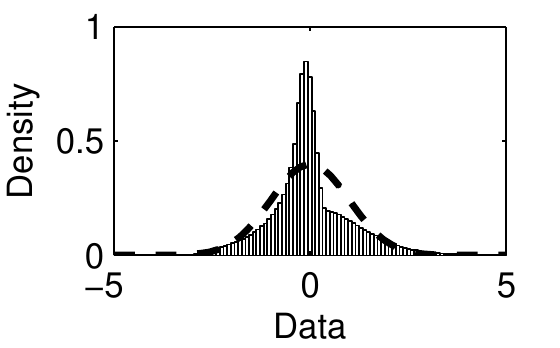}}
	\caption{Histograms of normalized displacement $\bar{Z}(t)$ with standard normal density based on 1000 MC realizations.}
	\label{3fig4}
\end{figure}

Let us briefly analyze the response of a harmonic oscillator with unit mass forced by jumping process $Y(t)$ with $\bar{f}=2.67$ Hz, employing MC to justify the normality assumptions. The coefficients of skewness and kurtosis of the state vector $\bs{X}(t)=[Z(t),\dot{Z}(t)]^T$ as functions of the oscillator eigenfrequency $\mbox{f}_1$ for two different values of viscous damping $\zeta$ are depicted in figure \ref{3fig3}.
Note that for frequency range $0.5-7$ Hz the response is approximately Gaussian. As was expected, worse convergence is achieved for higher damping values, \textit{cf} the Rosenblatt theorem \citep{Grigoriu_nong}. Normalized histograms of displacement for eigenfrequencies $\mbox{f}_1=4$ and $12$~Hz and both damping values are depicted in figure \ref{3fig4}. Other techniques can be applied for approximations of the response outside this frequency range, e.g. memoryless transformations of Brownian colored noise, but this lies beyond the scope of our paper. Based on heuristic arguments and the Central Limit Theorem, we can assume that the higher the number of active spectators, and the more complex the grandstand geometry is, the more Gaussian the response will be.

A spectral density approximation of the forcing process for the frequency domain solution, figure \ref{3fig2} (b), cannot be further simplified. This is because the FRF of the structure has sharp peaks, and thus exact function values are needed. Any approximation employing indicator functions in the vicinities of significant harmonics preserving variance would be inaccurate.

\begin{figure}
	\centering
	\subfloat[spectral density]{\includegraphics[scale=0.7]{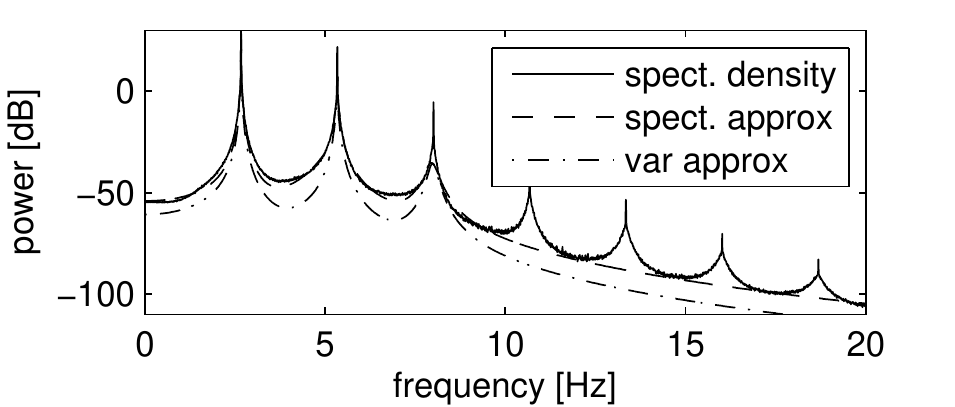}}
  \subfloat[spectral distribution function]{\includegraphics[scale=0.7]{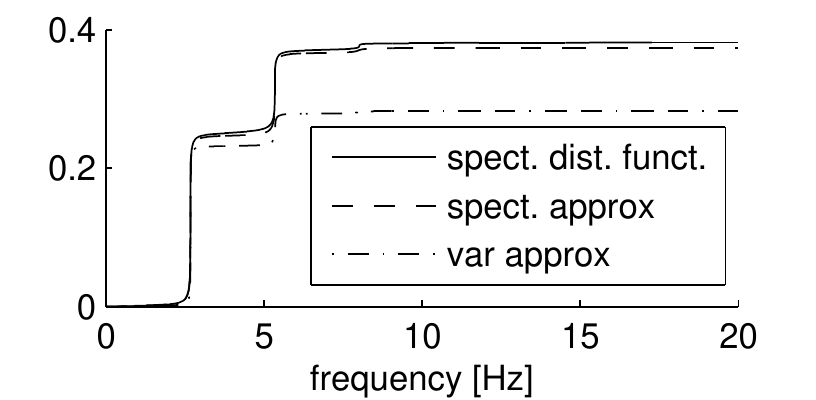}}
	\caption{Spectral density and spectral distribution function of centered forcing $\tilde{Y}(t)$ and of its approximation $\sum_{i=1}^6\hat{Y}_i(t)$, where $\hat{Y}_i(t)$ are independent $AR(2)$ processes with the coefficients in table~\ref{3tab2} based on spectral and variance optimization.}
	\label{3fig5}
\end{figure}

However, we can employ filtered white noise processes $AR(2)$, which arise as a solution of the second order It\^{o} equation
\begin{equation}
c_{2,i}\ddot{\hat{Y}}_i(t)+c_{3,i}\dot{\hat{Y}}_i(t)+c_{1,i}\hat{Y}_i(t)=W(t)
\label{3eq3}
\end{equation}
with spectral density
\begin{equation}
s_i(\omega)=\frac{1}{[c_{1,i}-c_{2,i}\omega^2]^2+(\omega c_{3,i})^2}
\label{3eq4}
\end{equation}
where $(c_{1,i},c_{2,i},c_{3,i})$ correspond to the stiffness, mass and damping of a harmonic oscillator. This function has a sharp peak positioned at $\mbox{f}_1=\sqrt{c_1/c_2}/2\pi$ when we neglect shifts due to damping effects. For a closer approximation, we assume $\tilde{Y}(t)\approx\sum_{i=1}^n\hat{Y}_i(t)$, where $\hat{Y}_i(t)$ are mutually independent $AR(2)$ processes. Identification leads to a nonlinear optimization problem: find such $c_{k,i}$, $k=1,2,3$ and $i=1,\dots,n$, that minimize $L_2$ norm $||\bullet||_{L_2}$ of the difference 
\begin{eqnarray}
&&e(\omega,\bs{c})=\hat{f}_{\tilde{Y}}(\omega)-\sum_{i=1}^n\frac{1}{[c_{1,i}-c_{2,i}\omega^2]^2+(\omega c_{3,i})^2}\label{3eq5}\\
&&\min_{\boldsymbol{c}=\{c_{1,1},\dots,c_{3,n}\}}||e(\omega,\bs{c})||_{L_2},\,\omega\in A\subset \mathbb{R}^+\mbox{ compact},\label{3eq6}
\end{eqnarray}
where $\hat{f}_{\tilde{Y}}$ denotes a spectral density estimate of the centered force term $\tilde{Y}(t)$. The problem can be solved by the Nonlinear Least Squares method, by Simulated Annealing etc, with easily estimated initial vector $\bs{c}_0$. Optimized coefficients for $n=6$ of the centered process $\tilde{Y}(t)$ are presented in table \ref{3tab2}, and the corresponding spectral density and spectral distribution function are presented in figure \ref{3fig5}. Note that the spectral density is two-sided, and only one half was integrated in the spectral distribution function, thus the variance indicated is $0.7627/2=0.3814$.

The performance of the structure was briefly quantified in section \ref{upcrossing}, where the two expressions (\ref{2eq22}) and (\ref{2eq24}) depended on the response variance. Thus an alternative approach is to optimize the response variance directly across some eigenfrequency range of a harmonic oscillator. Such coefficients are summarized in table \ref{3tab2}, with indices $\mathsf{var}$, spectral density and distribution function are depicted in figure \ref{3fig5}, frequency range $0.5-10$ Hz.
\begin{table}
	\centering
	\caption{Coefficients $c_{k,i}$, $k=1,2,3$ and $i=1,\dots,6$ of the six independent $AR(2)$ members used for approximation of the centered forcing term $\tilde{Y}(t)$ in frequency range $0.5-10$ Hz.}
	\begin{tabular}{|c|c|c|c|c|}\hline
	$i$ & $c_{1,i}$ & $c_{2,i}$ & $c_{3,i}$ & $\sqrt{c_{1,i}/c_{2,i}}/2\pi$ \\\hline
	1 & 90.8657 & 0.3227 & 0.0148 & 2.67 \\
	2 & 35.9464 & 0.1276 & 0.1167 & 2.67 \\
	3 & 283.6701 & 0.2520 & 0.0118 & 5.34 \\
	4 & 74.1544 & 0.0661 & 0.0737 & 5.33 \\
	5 & 913.9890 & 1.1120 & 21.5186 & 4.56 \\
	6 & 228.5270 & 0.0907 & 0.1576 & 7.99 \\\hline
  1$_{\mathsf{var}}$ & 91.0909 & 0.3237 & 0.0076 & 2.67 \\
	2$_{\mathsf{var}}$ & 40.3066 & 0.1430 & 0.0804 & 2.67 \\
	3$_{\mathsf{var}}$ & 281.0462 & 0.2490 & 0.0209 & 5.35 \\
	4$_{\mathsf{var}}$ & 83.7399 & 0.0746 & 0.0573 & 5.33 \\
	5$_{\mathsf{var}}$ & 914.0015 & 0.9376 & 21.6921 & 4.97 \\
	6$_{\mathsf{var}}$ & 228.7642 & 0.0908 & 0.1552 & 7.99 \\\hline
	\end{tabular}
	\label{3tab2}
\end{table}

Let us briefly note the situation when the forcing process $Y(t)$ is not a unit process, i.e. $G_H\neq 1$. Since we are limiting our considerations to Gaussian approximation, only the first two moments of $G_H$ will apply. The deterministic weight corresponds to a singular case $\mathsf{var}G_H=0$. Then the forcing term has the form $Y_G(t)=G_HY(t)$, mean response $\mu_{Z,G}(t)=\mathsf{E}[G_H]\mu_Z(t)$ and stationary response variance $\sigma_{Z,G}^2=\mathsf{E}[G_H^2]\sigma_Z^2$, where $\mu_Z(t)$ and $\sigma_Z^2$ are the response mean and the variance of the structure loaded by the unit forcing term $Y(t)$. However we should be aware that even when process $Y(t)$ was a Gaussian, process $Y_G(t)$ as a product of a random variable with a stochastic process, is not Gaussian. To quantify the influence of such scaling, compare the histograms in figure \ref{3fig2} (c) and \ref{3fig2} (d), where $G_H$ has normal distribution with mean value $0.7709$ and variance $0.0167$.
%
\section{Numerical examples and comparison}
\label{examples}
In this section, the quality of the approximation in the time or frequency domain will be compared with MC simulation.
\begin{figure}
	\centering
	\begin{tabular}{ccc}
  \includegraphics[scale=0.6]{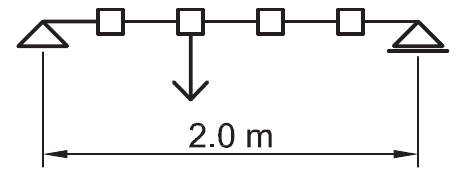} &
  \includegraphics[scale=0.3]{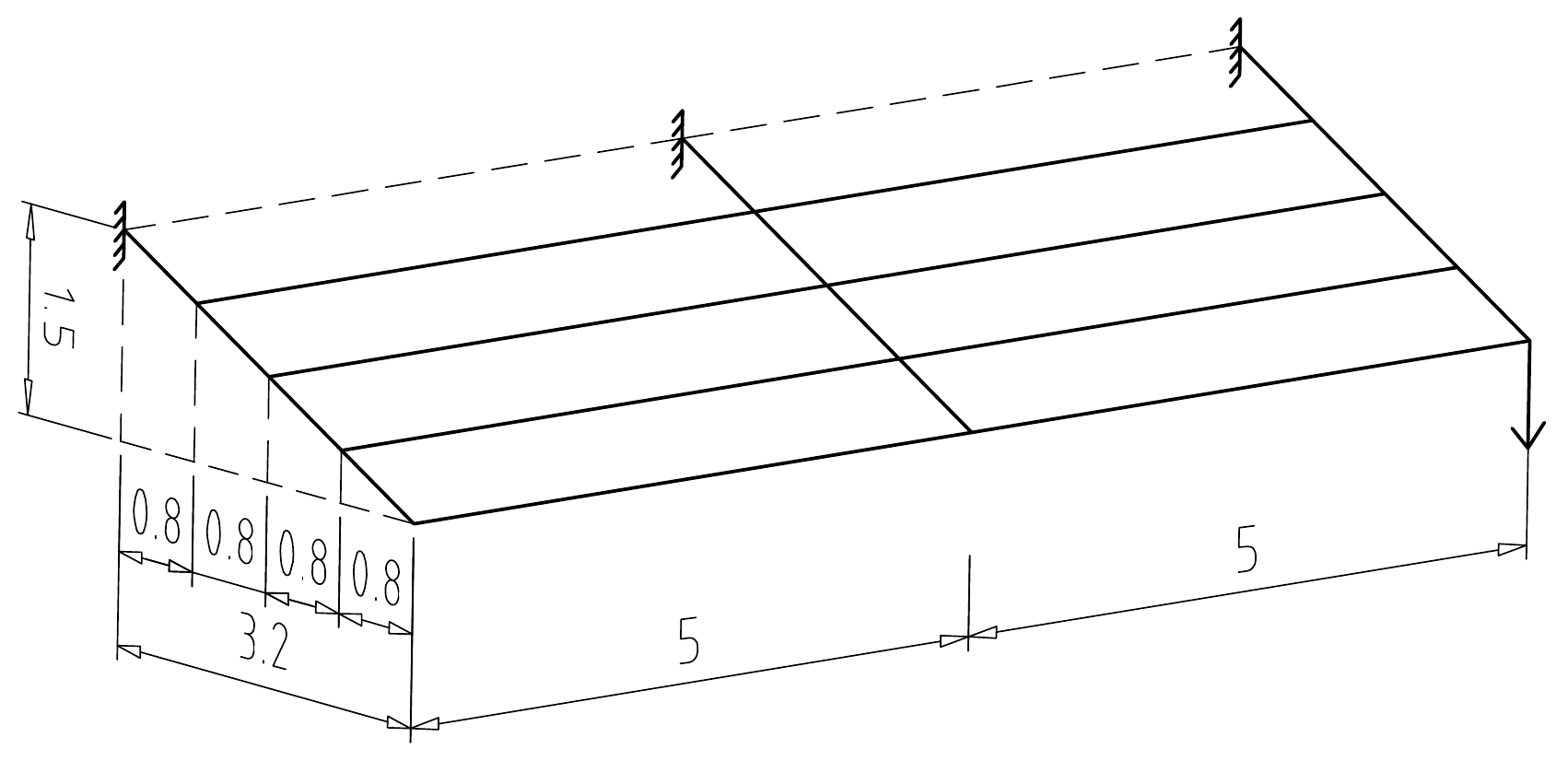} &
  \includegraphics[scale=0.4]{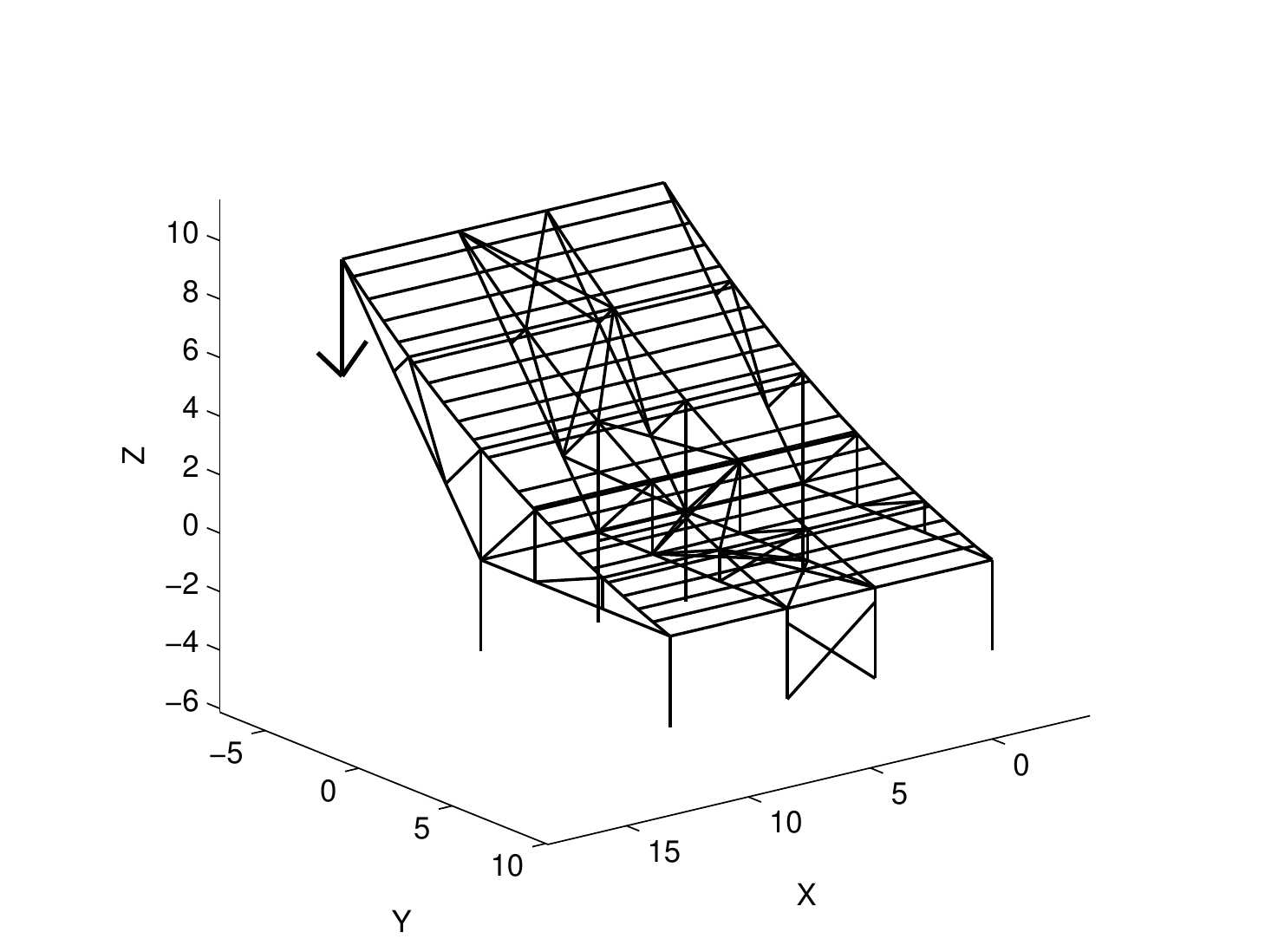} \\
  \includegraphics[scale=0.5]{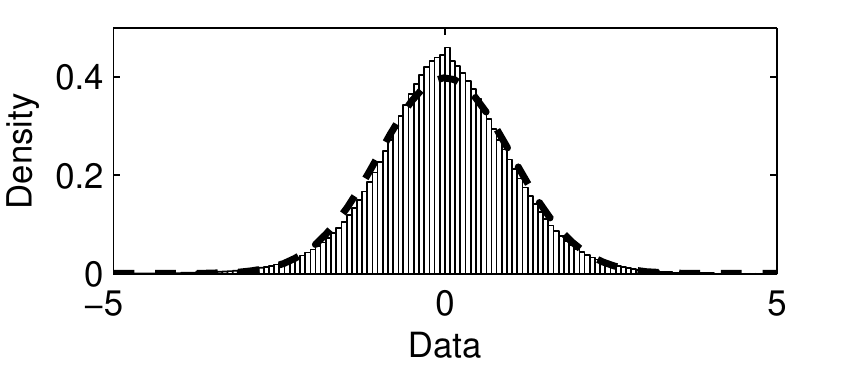} &
  \includegraphics[scale=0.5]{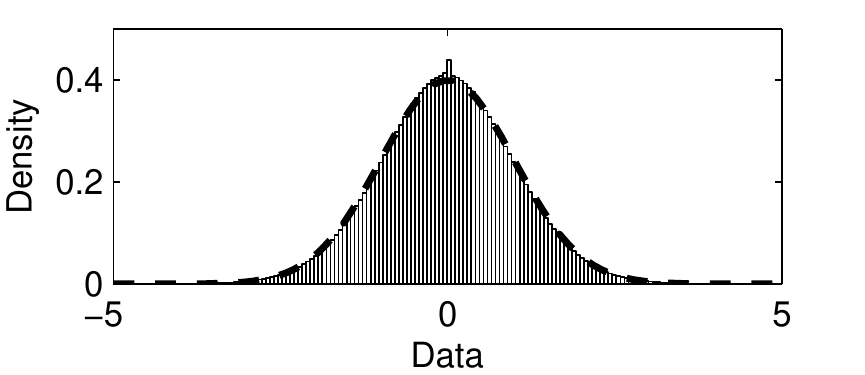} &
  \includegraphics[scale=0.5]{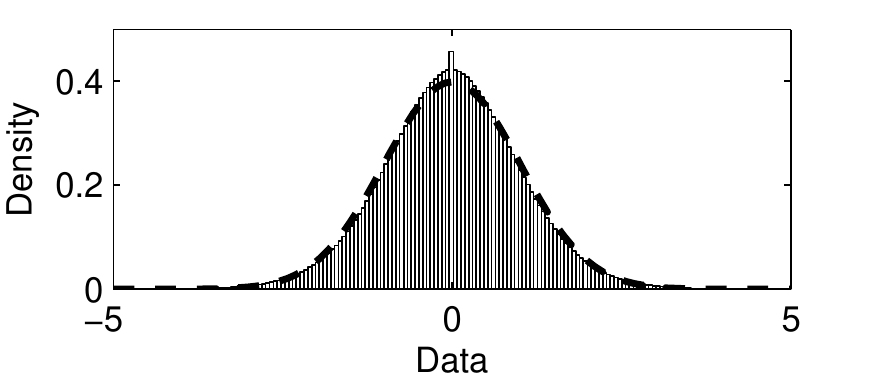}
  \end{tabular}
	\caption{Tested structures, geometries with labeled points of interest and normed histograms of the displacement, structures occupied by an active crowd only.}
	\label{4fig1}
\end{figure}
A total of four mechanical systems will be tested: a harmonic oscillator, a simply supported beam, a simple cantilever grandstand, and a realistic grandstand with 1, 4, 72 and 630 positions for spectators, respectively. Geometries with centered normed response histograms for an active crowd only, based on MC simulation, are depicted in figure \ref{4fig1}, and several lowest eigenfrequencies corresponding to the vertical bending modes are presented in table \ref{4tab1}. Note also that assumptions on convergence to normal distribution are approximately fulfilled. All examples are artificial, not realistic, so the measured responses would be unacceptable. 
\begin{table}
	\centering
	\caption{First eigenvalues corresponding to vertical bending modes [Hz], \textit{cf} figure \ref{4fig1}.}
	\begin{tabular}{|c|c|c|c|c|c|c|c|c|}
		\hline
		Structure & f$_1$ & f$_2$ & f$_3$ & f$_4$ & f$_5$  & f$_6$ & f$_7$ & f$_8$ \\\hline
		beam & 7.5 & 22.5 & 40.3 & --- & --- & --- & --- & --- \\
		cantilever & 5.4 & 7.0 & 8.2 & 25.1 & 28.4 & --- & --- & --- \\
		grandstand & 2.5 & 2.6 & 3.8 & 4.5 & 4.8 & 4.9 & 5.4 & 6.0 \\\hline
	\end{tabular}
	\label{4tab1}
\end{table}
%
\subsection{Harmonic oscillator}
\label{harmosc}
\begin{figure}
	\centering
	\includegraphics[scale=0.6]{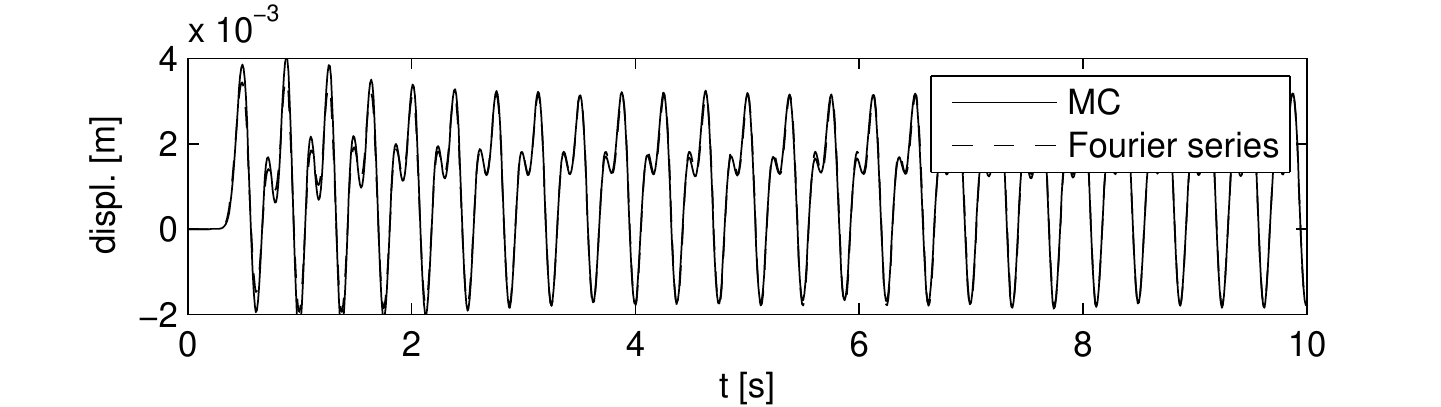}
	\caption{Response mean displacement in comparison with an approximation based on the first four harmonics for a harmonic oscillator, $f_1=5$ Hz $\zeta=0.07$.}
	\label{4fig2}
\end{figure}

\begin{figure}
	\centering
\subfloat[$\zeta=0.001$]{\includegraphics[scale=0.7]{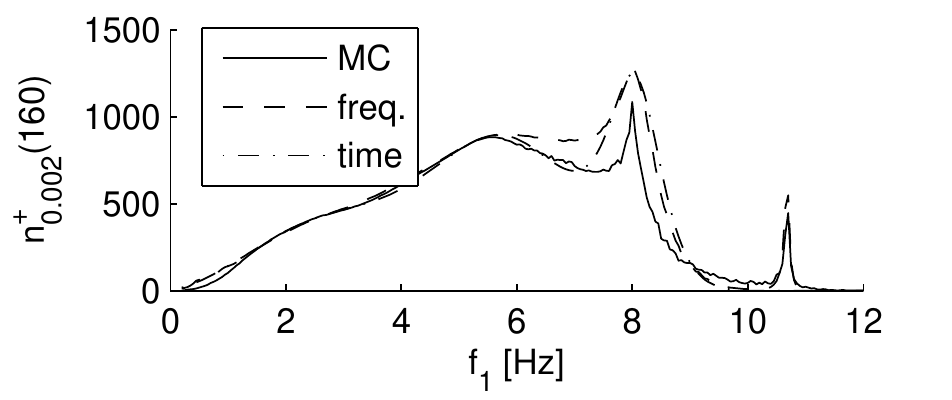}}
\subfloat[$\zeta=0.001$]{\includegraphics[scale=0.7]{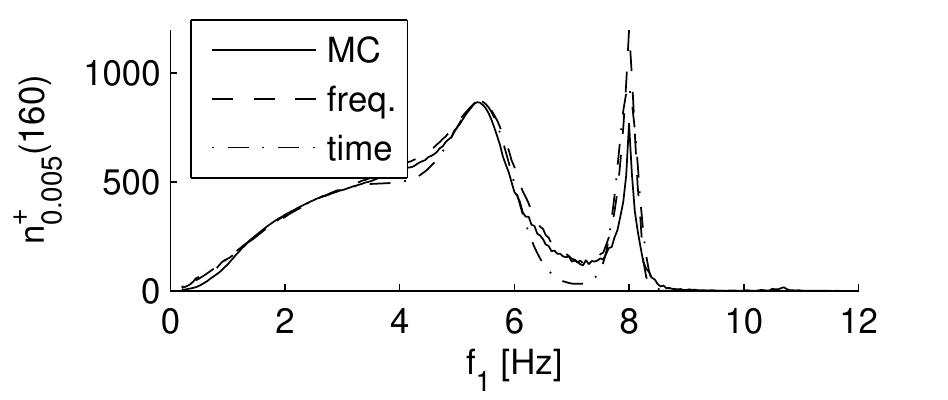}}\\
\subfloat[$\zeta=0.07$]{\includegraphics[scale=0.7]{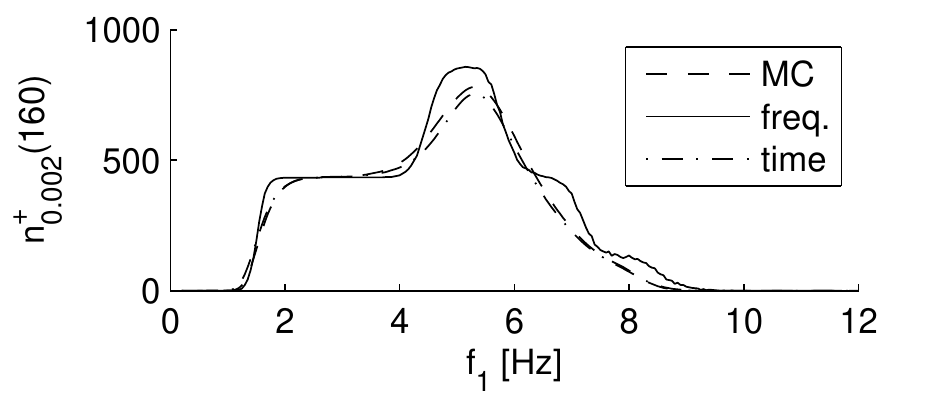}}
\subfloat[$\zeta=0.07$]{\includegraphics[scale=0.7]{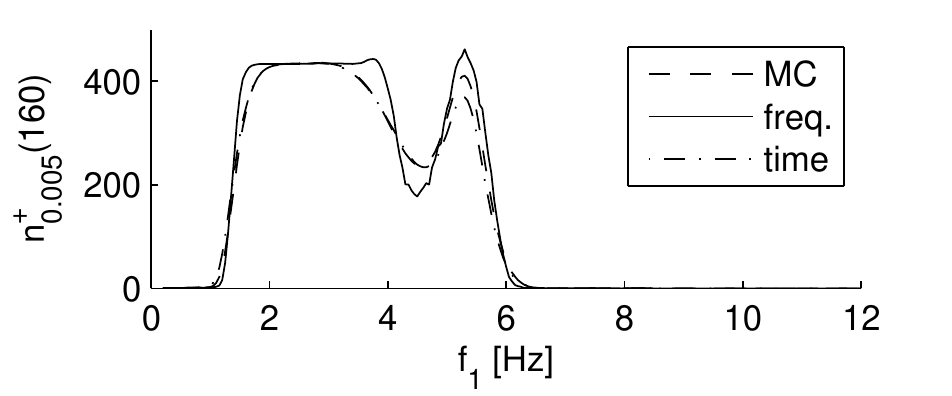}}\\
\subfloat[$\zeta=0.001$]{\includegraphics[scale=0.7]{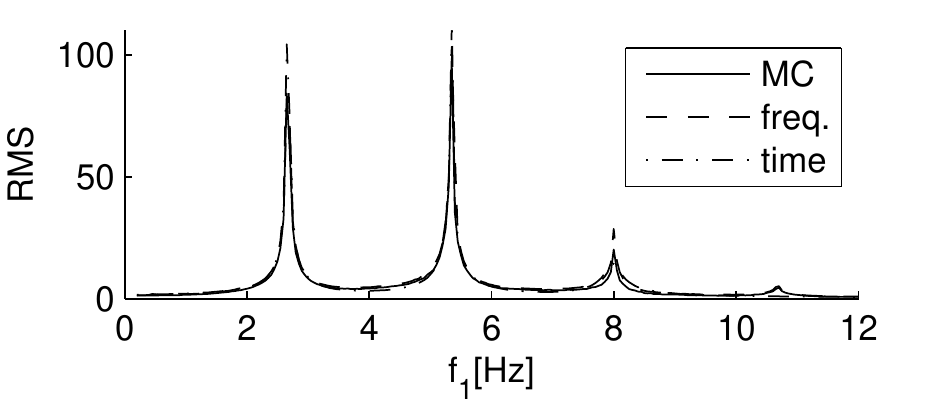}}
\subfloat[$\zeta=0.07$]{\includegraphics[scale=0.7]{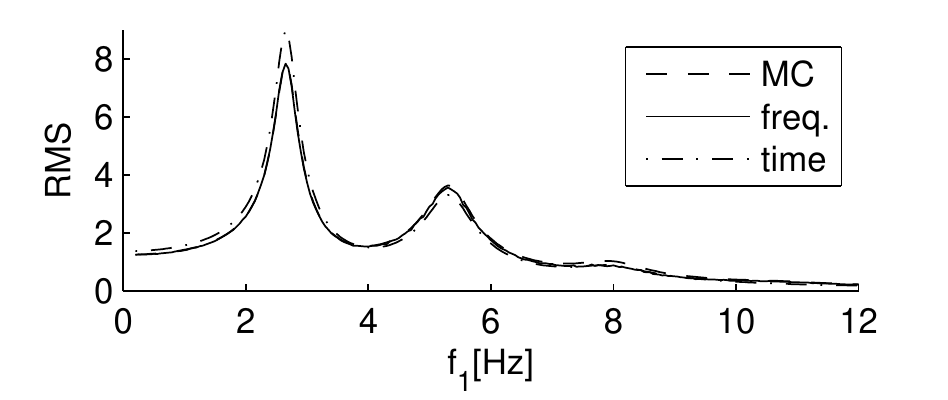}}
	\caption{Total mean up-crossings $n_x^+(T)$ and acceleration $RMS$ values as functions of $\mbox{f}_1$ for a harmonic oscillator, distinct levels $x$ and viscous damping $\zeta$, $T=160$ s.}
	\label{4fig3}
\end{figure}

Let us assume a harmonic oscillator with unit mass, two values of viscous damping $\zeta$ and variable stiffness. The mean value response is presented in figure \ref{4fig2}. The response was acquired by direct integration of (\ref{2eq2}), starting at $t=0.5$~s. The approximation utilizes equation (\ref{3eq1}) and the coefficients in table \ref{3tab1}. A comparison of the total mean up-crossings $n_x^+(T)$ in the time interval $[0,T]$, $T=160$~s, as functions of oscillator eigenfrequency $\mbox{f}_1$ for two values of viscous damping $\zeta=0.001$ and $\zeta=0.07$ and for two fixed levels $x=0.002$ and $x=0.005$~m are depicted in figure \ref{4fig3}, employing formulas (\ref{2eq22}) and (\ref{2eq23}). The size of time interval $T$ is based on heuristic considerations about the average length of the musical compositions. The time domain solution is based on the sum of six independent $AR(2)$ processes with the coefficients in table \ref{3tab2} for $\mathsf{var}$ optimization. The stationary response variances are computed according to formulas (\ref{modal5}), (\ref{modal6}) and (\ref{2eq14a}). The frequency domain approximation employs equations (\ref{2eq19}), (\ref{2eq20}) and (\ref{2eq21}). The spectral density estimate $\hat{f}_{\tilde{Y}}(\omega)$ of the centered input process has $307$ values over the frequency range $0-10$~Hz, using variable division. Figure \ref{4fig3} shows that the results are roughly in agreement with MC in the frequency range $0.2-10$~Hz. Note that the total number of mean zero-up-crossings for a deterministic periodic function with frequency $2.67$ Hz is $160\cdot 2.67\approx 427$, \textit{cf} figure \ref{4fig3} (c) and (d), where distinct plateaux are found. The results for MC are based on $1000$ realizations $160$ s in length.
%
\subsection{A simply supported beam}
\label{beam}
\begin{figure}
	\centering
\subfloat[active crowd only]{\includegraphics[scale=0.7]{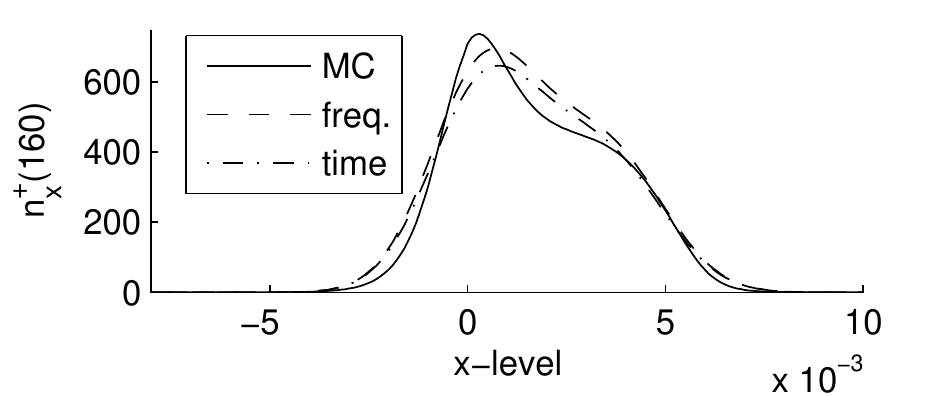}}
\subfloat[2 active, 2 passive spectators]{\includegraphics[scale=0.7]{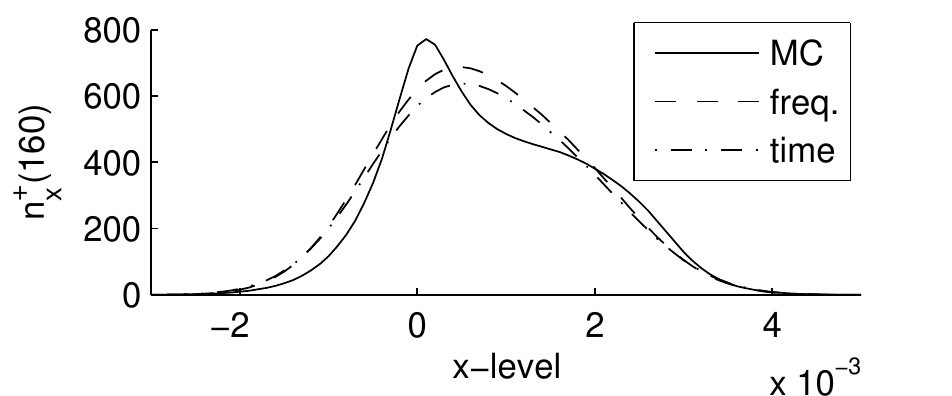}}\\
\subfloat[single realization]{\includegraphics[scale=0.7]{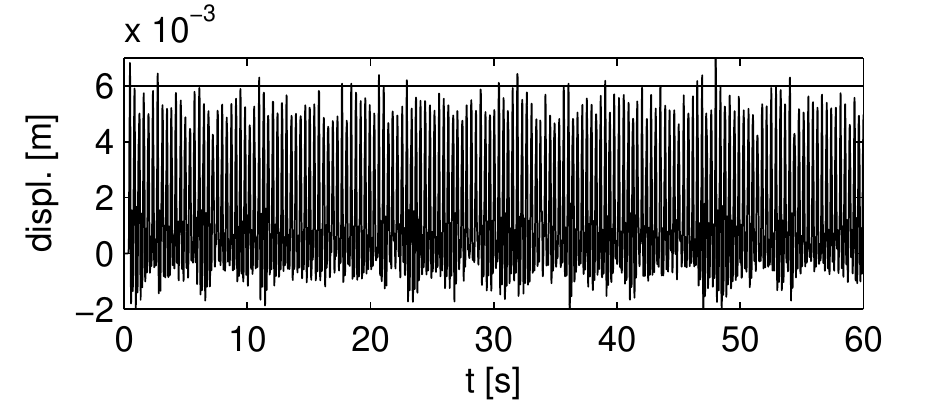}}
\subfloat[response mean]{\includegraphics[scale=0.7]{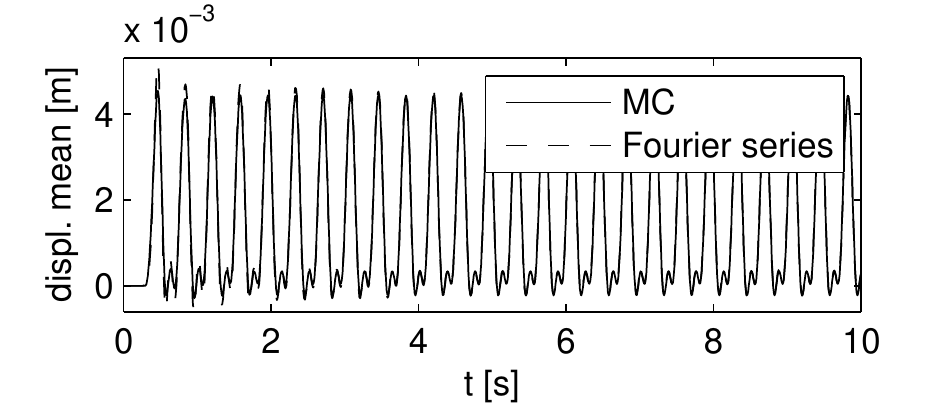}}
	\caption{Total mean up-crossings $n_x^+(T)$ of a simply supported beam as functions of $x$, $T=160$~s for an active crowd (a) and for a mixed crowd (b), single realization for an active crowd (c) and mean response for an active crowd (d).}
	\label{4fig4}
\end{figure}

The next example is a simply supported beam with Rayleigh damping $\zeta_1=0.05$ and $\zeta_2=0.08$ for the first two vertical modes and total mass $1700$~kg. Poor approximations are anticipated, since the structure is quite stiff with high eigenfrequencies, see table \ref{4tab1} and the results for the harmonic oscillator. Two cases are studied: a structure occupied by an active crowd only; and a structure occupied by a mixed crowd. In the second case, the two left hand side positions are loaded by forces and the two right hand side positions are occupied by passive spectators. Deterministic biodynamic models according to Coermann are used. These are simple-degree-of-freedom oscillators with mass $86.2$~kg, stiffness $85.25$~kN/m and viscous damping $1.72$~kNs/m, eigenfrequency $5$~Hz, \textit{cf} \citep{Sachse}. Results are presented only for the labeled point in figure \ref{4fig1}. The total mean up-crossings for the first case $n_x^+(160)$ as a function of level $x$ are depicted in figure \ref{4fig4} (a). The $RMS$ values for acceleration are $1.947$~m/s$^2$ for the MC solution, $2.135$~m/s$^2$ for the frequency domain solution, and $1.914$~m/s$^2$ for the time domain solution. All input processes are treated as independent, so matrix $\bs{S}_{\tilde{Y}\tilde{Y}}$ in equation (\ref{2eq19}) has nonzero only diagonal entries. The mean value response is depicted in figure \ref{4fig4} (d) with a single realization (c). The total up-crossings for a mixed crowd are depicted in figure \ref{4fig4} (b), and the $RMS$ values are $1.048$~m/s$^2$ for the MC solution, $1.094$~m/s$^2$ for the frequency domain solution, and $0.990$~m/s$^2$ for the time domain solution. Let us also recall our assumption of fixed spatial distribution of the crowd, mass coefficient $\gamma=m_H/m_S=0.1$, where $m_H$ denotes the total mass of passive spectators, and $m_S$ denotes the total mass of the structure. Results for MC based on $2000$ realizations.
The approximate shape of the total up-crossings, especially for a mixed crowd, differ from the MC simulation, because of the non-Gaussian response due to high structure eigenfrequencies.
%
\subsection{Cantilever grandstand}
\label{cantilever}
\begin{figure}
	\centering
\subfloat[active crowd only]{\includegraphics[scale=0.7]{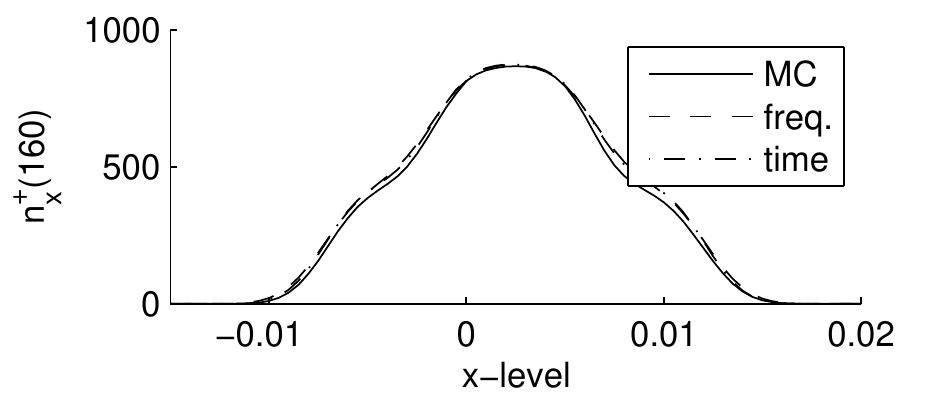}}
\subfloat[36 active, 36 passive spectators]{\includegraphics[scale=0.7]{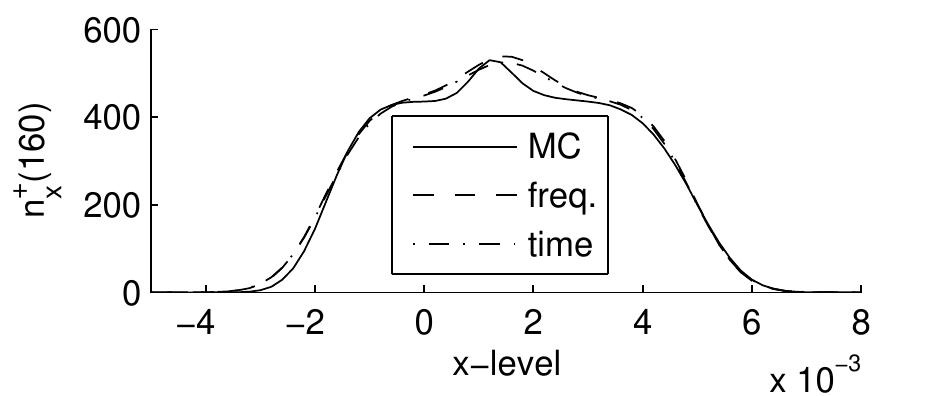}}\\
\subfloat[single realization]{\includegraphics[scale=0.7]{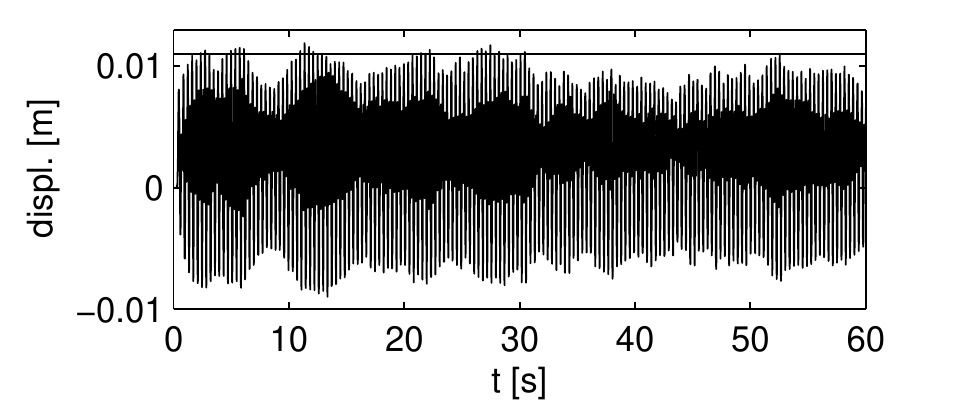}}
\subfloat[response mean]{\includegraphics[scale=0.7]{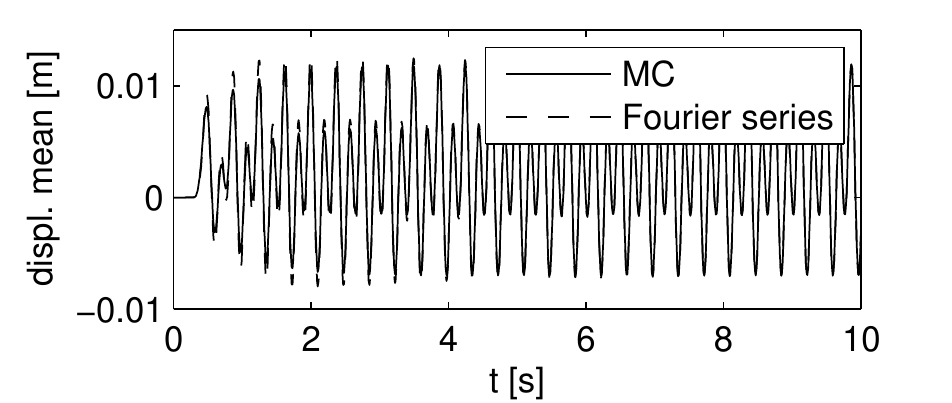}}
	\caption{Total mean up-crossings of a cantilever grandstand, active crowd (a) and mixed crowd (b), single realization (c) and mean response (d) for an active crowd.}
	\label{4fig5}
\end{figure}

This system has total mass $18.2$~t, geometry according to figure \ref{4fig1}, and is loaded with 72 active spectators in the first case, Rayleigh damping with $\zeta_1=0.05$ and $\zeta_2=0.08$ is used for the first two vertical modes. The total up-crossings of the response displacement are depicted in figure \ref{4fig5} (a), and single realization and the mean response are depicted in sub-figures (c) and (d), $RMS$ accelerations $5.583$~m/s$^2$ for the MC solution, $5.682$~m/s$^2$ for the frequency domain solution, and $5.616$~m/s$^2$ for the time domain solution. For 36 spectators chosen to be passive according to Coermann with uniformly random but fixed positions, the resulting up-crossings are depicted in sub-figure (b), mass coefficient $\gamma=0.17$. The acceleration $RMS$ values in this case appear to be $1.327$~m/s$^2$ for the MC solution, $1.347$~m/s$^2$ for the frequency domain solution, and $1.309$~m/s$^2$ for the time domain solution. The results for MC are again based on $2000$ realizations $160$ s in length.
%
\subsection{Realistic grandstand}
\label{grandstand}
\begin{figure}
	\centering
\subfloat[active crowd only]{\includegraphics[scale=0.7]{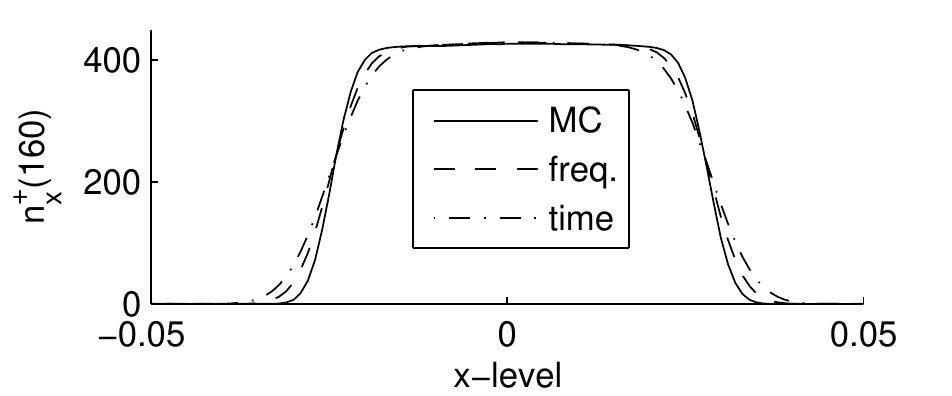}}
\subfloat[315 active, 315 passive spectators]{\includegraphics[scale=0.7]{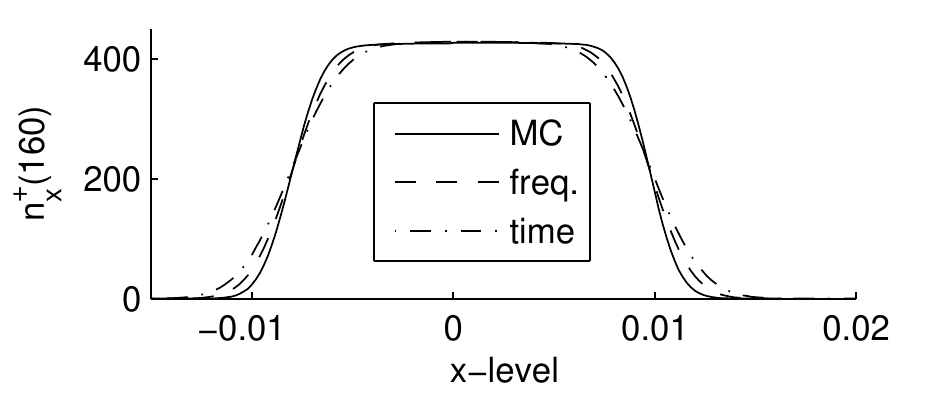}}\\
\subfloat[single realization]{\includegraphics[scale=0.7]{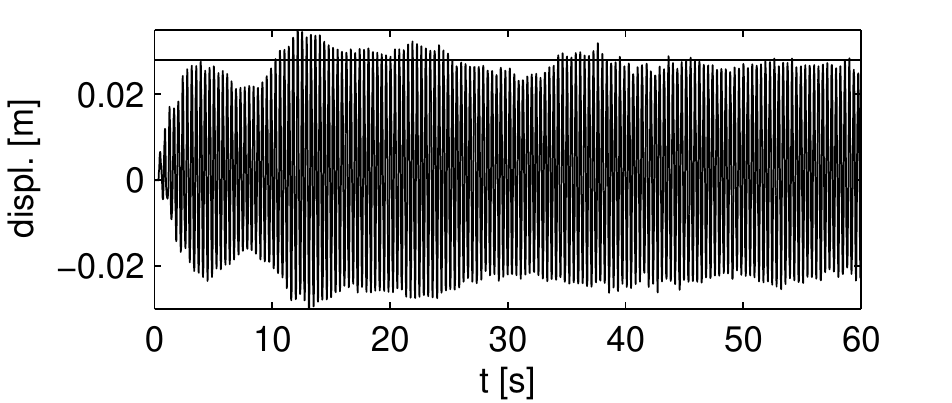}}
\subfloat[response mean]{\includegraphics[scale=0.7]{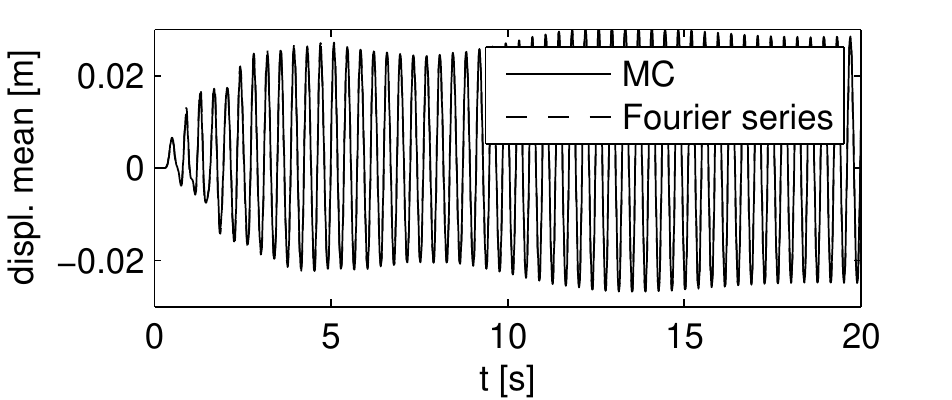}}
	\caption{Total mean up-crossings of a realistic grandstand, active crowd (a), mixed crowd (b), single realization (c) and mean response (d) for an active crowd.}
	\label{4fig6}
\end{figure}

In this concluding example, let us briefly examine the results acquired for a complex structure. The geometry is sketched in figure \ref{4fig1} with the point of interest labeled, total mass $148.6$~t, Rayleigh damping with $\zeta_1=0.01$ for the first and $\zeta_2=0.02$ for the sixth vertical mode used here. The results for the structure loaded by an active crowd only are depicted in figure \ref{4fig6} (a) (c) (d). The acceleration $RMS$ values appear to be $5.546$~m/s$^2$ for the MC solution, $5.565$~m/s$^2$ for the frequency domain solution, and $5.597$~m/s$^2$ for the time domain solution. In the second case, $315$ positions are occupied by passive spectators according to Coermann, mass coefficient $\gamma=0.18$. The results for this case are presented in figure \ref{4fig6} (b), $RMS$ accelerations $1.905$~m/s$^2$ for the MC solution, $1.904$~m/s$^2$ for the frequency domain solution, and $1.919$~m/s$^2$ for the time domain solution. MC simulation based on $2000$ realizations.
%
\subsection{Comparison and performance}
\label{performance}
\begin{figure}
	\centering
	\subfloat[$x=0.005$]{\includegraphics[scale=0.7]{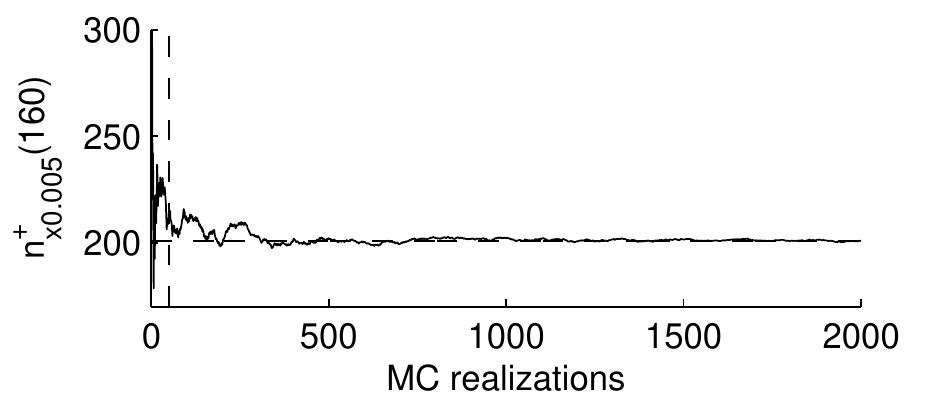}}
	\subfloat[$x=0.006$]{\includegraphics[scale=0.7]{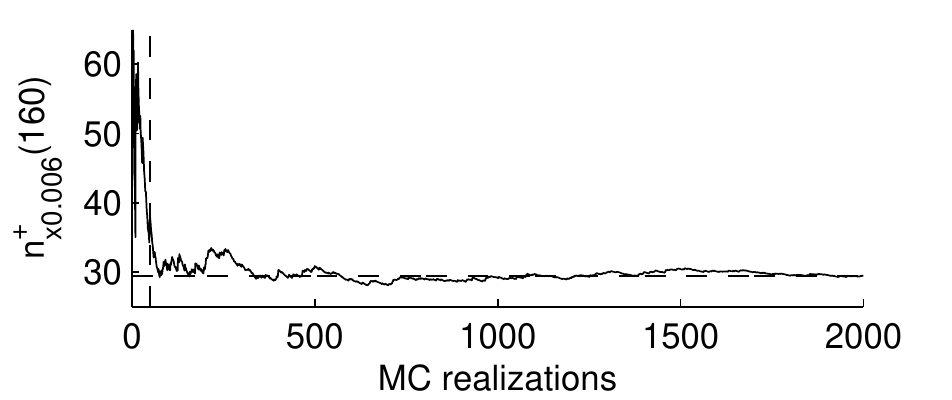}}
	\caption{MC convergence tests of $n_x^+(160)$ for a cantilever grandstand, \textit{cf} figure \ref{4fig5} (b).}
\label{4fig7}
\end{figure}

The time consumption for the different solution techniques is summarized in table \ref{4tab2}, where the demands of MC simulation are presented for 50 realizations only. This value is based on the convergence tests of total up-crossings presented in figures \ref{4fig7} (a) and (b) for a cantilever grandstand. Obviously these tests are highly sensitive to level $x$, and the  value used can be considered as a lower bound. The size of the time integration step is chosen to be $h=0.01$~s, Newmark integration scheme used. The number of dofs of each system is also mentioned, together with the size of the Lyapunov equation (\ref{2eq14}) $n_\mathrm{L}$ for a mixed crowd. Obviously $n_\mathrm{L}=2n_{\mathrm{dof}}+2n_{\mathrm{p}}+12n_{\mathrm{a}}$ where $n_{\mathrm{dof}}$ denotes the number of dofs of the empty structure, $n_{\mathrm{p}}$ is the number of dofs of the passive spectators, and $n_{\mathrm{a}}$ is the number of active spectators. In the case of a cantilever grandstand we therefore have $2\cdot 504+2\cdot 36+12\cdot 36=1512$, since passive spectators are modeled as systems with a single degree of freedom. For the purposes of comparison, the time consumption for the partial modal transform are also summarized together with the number of eigenvectors used $n_{\mathrm{eig}}$ and according highest frequency f$_{\mathrm{eig}}$. All simulations were performed on core i7 with a 16~GB RAM computer, Matlab\textsuperscript{\textregistered} parallel implementation.
\begin{table}
\centering
\caption{Comparison of computational demands.}
\begin{tabular}{|c|r|r|r|r|}\hline
	Method/Struct. & \multicolumn{1}{c|}{Harm. osc.} & \multicolumn{1}{c|}{Beam} & \multicolumn{1}{c|}{Cant. grand.} & \multicolumn{1}{c|}{Real. grand.} \\\hline
	\multicolumn{5}{|c|}{full system} \\\hline
  $n_{\mathrm{dof}}$ & 1 & 29 & 504 & 4\ 068 \\
	$n_\mathrm{L}$ & 14 & 86 & 1512 & 12\ 546 \\	
	MC 50 & 0.601 s & 20.078 s & 380.451 s & 5\ 484 s \\
	Freq. domain & 0.106 s & 2.672 s & 74.639 s & 4\ 178 s \\
	Time domain & 0.112 s & 2.392 s & 38.915 s & 6\ 507 s \\\hline
	\multicolumn{5}{|c|}{partial modal transform, \textit{cf} equations (\ref{modal4}), (\ref{modal5}) and (\ref{modal6})} \\\hline
	$n_{\mathrm{eig}}$ & -- & 7 & 10 & 50 \\
	f$_{\mathrm{eig}}$ & -- & $37$ Hz & $30$ Hz & $23$ Hz \\
	$n_\mathrm{L}$ & -- & 42 & 524 & 4\ 510 \\
	MC 50 & -- & 13.240 s & 36.214 s & 307 s \\
	Freq. domain & -- & 1.764 s & 4.452 s & 734 s \\
	Time domain & -- & 1.588 s & 2.582 s & 67 s \\\hline
\end{tabular}
\label{4tab2}
\end{table}
The results in the table suggest that when 50 realizations are employed, only structures of moderate size can be effectively solved by semi-analytical methods.
%
\section{Conclusions}
\label{concl}
This paper has presented a study of the vibration of grandstands loaded by an active crowd, using Gaussian approximation of the response. The main results can be summarized as follows:
\begin{enumerate}
	\item A mathematical description of the response of a mechanical system employing spectral and time domain solutions for weakly stationary Gaussian excitations has been recalled. Partial modal transformation due to a passive crowd has been briefly discussed.
	\item A motivating example of a harmonic oscillator has shown that the normalized displacement and velocity have approximately normal distribution under the conditions on eigenfrequencies and damping.
	\item Taking this fact into account, the mean value of the force has been approximated as a truncated Fourier series, and the spectral density of the centered process has been estimated employing the Parzen window for the frequency domain solution. For the time domain solution, we have employed a linear combination of independent auto-regression processes of the second order with coefficients optimized to achieve the least error in the response variance of a harmonic oscillator.
	\item Three different examples of varying complexity have shown the quality of the response approximation in terms of total displacement up-crossings and acceleration $RMS$ in comparison with Monte Carlo simulation. Limitations following from a simple oscillator have been confirmed on multi-degree-of-freedom systems. 
	\item The computational demands have been measured and summarized in terms of the time needed for solution, and the applicability of the techniques has been proved.
\end{enumerate}
Finally, let us note that these methods are approximations and can be further refined. In particular, the solution can be reformulated for the non-Gaussian processes for which higher moments can be derived. These apply in improving the mean up-crossing rate estimates for stiff structures.
%
\section*{Acknowledgement}
Financial support for this work from the Czech Technical University in Prague
under project No. SGS12/027/OHK1/1T/11 and from the Czech Science Foundation under project No. GAP105/11/1529 is gratefully acknowledged.



\end{document}